\documentclass[a4paper,fleqn,usenatbib]{mnras}

\usepackage{newtxtext,newtxmath}

\usepackage[T1]{fontenc}
\usepackage{ae,aecompl}

\usepackage{graphicx}	
\usepackage{amsmath}	
\usepackage{amssymb}	

\usepackage{supertabular}
\usepackage{caption}

\usepackage{booktabs}

\title[Cosmic rays from multifrequency data]{Imprints of Cosmic Rays in Multifrequency Observations of the Interstellar Emission}

\author[E. Orlando]{
E. Orlando$^{1}$\thanks{E-mail: eorlando@stanford.edu}
\\
$^{1}$Hansen Experimental Physics Laboratory and Kavli Institute for Particle Astrophysics and Cosmology, Stanford University, Stanford, CA 94305, USA\\
}

\date{Accepted XXX. Received YYY; in original form ZZZ}

\pubyear{2015}

\begin{document}
\label{firstpage}
\pagerange{\pageref{firstpage}--\pageref{lastpage}}
\maketitle

\begin{abstract}
Ever since the discovery of Cosmic Rays (CRs), significant advancements have been made in modeling their propagation in the Galaxy and in the Heliosphere. However, propagation models suffer from degeneracy of many parameters. To complicate the picture the precision of recent data have started challenging existing models. \\
To tackle these issues we use available multifrequency observations of the interstellar emission from radio to gamma rays, together with direct CR measurements, to study local interstellar spectra (LIS) and propagation models. \\
As a result, the electron LIS is characterized without any assumption on  solar modulation, and favorite propagation models are put forward. 
More precisely, our analysis leads to the following main conclusions: (1) the electron injection spectrum needs at least a break below a few GeV; (2) even though consistent with direct CR measurements, propagation models producing a LIS with large all-electron density from a few hundreds of MeV to a few GeV are disfavored by both radio and gamma-ray observations; (3) the usual assumption that direct CR measurements, after accounting for solar modulation, are representative of the proton LIS in our  $\sim$1~kpc region is challenged by the observed local gamma-ray HI emissivity. \\
We provide the resulting proton LIS, all-electron LIS, and propagation parameters, based on synchrotron, gamma-ray, and direct CR data. A plain diffusion model and a tentative diffusive-reacceleration model are put forward. The various models are investigated in the inner-Galaxy region in X-rays and gamma rays. Predictions of the interstellar emission for future gamma-ray instruments ({\em e-ASTROGAM} and {\em AMEGO}) are derived. 
\end{abstract}




\section{Introduction}

The Milky Way is permeated by Cosmic Rays (CRs) that diffuse and interact within the Galaxy producing diffuse interstellar emission
from radio to gamma rays. While significant advancements have been made by studying CRs
through their diffuse interstellar emission either at radio \citep[e.g.][]{Strong2011} or at gamma-ray energies \citep[e.g.][]{diffuse1, diffuse2} independently, these studies are unavoidably affected
by uncertainties.
However, the CRs responsible for the radio emission are the same producing also the gamma-ray emission. In this work
we take advantage of this property with the aim of constraining CRs by looking at the interstellar emission in radio and gamma-ray energies simultaneously. This approach
provides a handle on both sides of the electromagnetic spectrum in understanding CRs, 
thereby leaving less room to uncertainties. 
Our very first attempt with this work shows that this approach is feasible. \\
 
In more detail,  
many studies on CR Local Interstellar Spectra\footnote{We define the LIS as the spectra of CRs in the local interstellar medium (within about 1~kpc from the Sun).} (LIS) and CR propagation models in the Galaxy have been performed thanks to sophisticated propagation codes \citep[e.g.][]{Moskalenko2015,Boschini, Dragon, Picard, Usine} and to unprecedented precise CR measurements. 
Even though the main interaction processes are identified, 
details on CR propagation models, on injection spectra in the interstellar medium,
and on the LIS are still uncertain.  
Some recent direct measurements of CRs are provided by PAMELA \citep{PAMELA} launched in 2006, by the Fermi Large Area
Telescope \citep[LAT,][]{Atwood} in orbit since 2008, and by the Alpha Magnetic Spectrometer-02 \citep[AMS-02,][]{AMS02}
working since 2011. These instruments have greatly reduced statistical and systematic uncertainties in measuring
CR fluxes, and are challenging present propagation models \citep[e.g.][]{Adriani2009} . Very recent {\em Fermi} electron measurements \citep{Fermiele} are found in agreement with AMS-02 data. 
Further recent CR measurements by \cite{Voyager} are performed with {\em Voyager~1} \citep{Stone}. 
Launched in 1977, {\em Voyager~1} has reached
interstellar space, providing measurements of CRs beyond the
influence of the solar modulation. CR measurements have enabled important studies \citep[e.g.][]{Moskalenko2002,Donato,Maurin,Donato2011,Aloisio,Tomassetti} by using a
well-established method based on comparing CR propagation models to CRs measurements \citep[e.g.][]{Gulli, Gaggero, Boschini, Dragon}. \\ 
CR all-electrons (electrons plus positrons), protons, and heavier nuclei interact with the gas in the interstellar medium and with the interstellar radiation field (ISRF) producing gamma rays via bremsstrahlung, inverse Compton (IC) scattering, and pion decay. 
The same CR all-electrons spiraling in the magnetic field produce synchrotron emission observed in radio and microwaves. 
The spectra of multiwavelenght observations of the interstellar emission reflects the spectra of CRs. In particular these multiwavelenght observations provide indirect CR measurements, which can extend beyond the local direct measurements and are not affected by solar modulation. Hence, they 
complement direct CR measurements for obtaining the LIS (defined in a region around 1~kpc from the Sun) and CR spectra throughout the Galaxy. 
Indeed, over the past years gamma-ray and radio/microwave observations of the interstellar emission have been used to gain information on CRs together with CR direct measurements and propagation models. However, this has been done performing gamma-ray and radio analyses separately.
More in detail, important studies on large-scale CRs and propagation models by observing the interstellar emission at gamma-ray energies have been performed since the 90's \citep[e.g.][]{Mori97, Pohl, Moskalenko98a}. Recently, a detailed work in \cite{diffuse2} investigated CR propagation models by studying the interstellar gamma-ray emission seen by {\em Fermi} LAT. The emission was computed for 128 propagation models: all the models provide a good agreement with gamma-ray data, but no best model was found, emphasizing the degeneracies among input parameters. Only standard reacceleration models were used.
At the opposite end of the electromagnetic spectrum, observations at the radio band of the interstellar synchrotron emission were used to constrain CRs and propagation models by \cite{Strong2011} finding that models with no reacceleration fit best synchrotron data.
This approach was followed by other similar works \citep[e.g.][]{Jaffe}. \cite{Orlando2013} investigated the spatial distribution of the synchrotron emission in temperature and polarization for the first time in the context of CR propagation models. Various CR source distributions, CR propagation halo sizes, propagation models (e.g. plain diffusion and diffusive-reacceleration models), and magnetic fields were tested against synchrotron observations, highlighting degeneracies among input parameters.  \\
As discussed by the previously referenced works, those studies suffer from unavoidable uncertainties and degeneracies given by the limited knowledge of many parameters (e.g. solar modulation, Galactic magnetic field, gas density, interstellar radiation field, propagation parameters, etc.) entering the modeling. 
To mitigate such uncertainties we study CRs properties by looking at the radio frequencies and gamma-ray energies simultaneously.
This allows for a handle on either side
of the electromagnetic spectrum steering the properties of the underlying CRs, thereby reducing degeneracies among the parameters. \\
More precisely, CR direct measurements below several GeV are usually used to derive the propagation parameters that are then applied
to the whole Galaxy \citep[e.g.][]{Gulli, Boschini, Dragon}. However, CR spectra below several GeV are affected by solar modulation \citep{Parker}, which leads to unavoidable approximations in the modeling. 
Below these energies the only available CR measurements that are unaffected by solar modulation are those from {\em Voyager~1}, which extend up to $\sim$70~MeV for all-electrons and up to a few hundreds of MeV/nucleon for hadrons only. 
As a consequence interstellar spectra in the energy range from 70 MeV to a few tens of GeV (for all-electrons) and a few hundreds of MeV/nucleon to a few tens of GeV/nucleon (for hadrons) are not directly measured by any instruments. Hence, usually in these ranges the LIS are obtained by interpolation and/or propagation models. In turn this range is very important for distinguishing CR propagation models in the entire Galaxy. \\ 
{\em In this work we use available spectral observations of the local gamma-ray emissivity and of the synchrotron emission, together with CR direct measurements to probe the CR LIS, and to specify preferred CR propagation models.}\\
We first introduce the method (Section 2) with the description of the models (Section 2.1) and the observations (Section 2.2). Results by comparing data and models are described in Section 3 and further used for predictions for future MeV missions in Section 4. In Section 5 we discuss the results and drive conclusions. 

\section{Method}

In the following we describe the general procedure adopted in this paper.
We start by using some latest available propagation models obtained with the GALPROP code, whose propagation parameters for hadrons are the result of our previous studies on recent CR measurements  
(details on the GALPROP code and on the propagation models used in this work are provided below).
Then, for each propagation model we infer the injection spectral parameters of primary electrons so that  
1) the propagated all-electron spectrum at Earth reproduces the CR measurements ({\em Voyager~1} and AMS-02 above a few ten GeV, which are unaffected by solar modulation), and 2) the calculated radio synchrotron emission reproduces the synchrotron spectral data as best as possible. 
In turn, the all-electron LIS is free from any approximation of the solar modulation effects, contrary to what is usually done.  The resulting all-electron spectra help in constraining propagation models, and also the proton LIS based on gamma-ray observations. \\
Details on the models follow in Section 2.1, while details on the data are given in Section 2.2.

\subsection{CR propagation models}
CR propagation models and associated interstellar emission are
built by using the GALPROP code\footnote{http://galprop.stanford.edu/}.

\subsubsection{Description of the GALPROP code}
The GALPROP code calculates the CR propagation in the Galaxy \citep[][and references therein]{Moskalenko98, Moskalenko2000, Strong2004, Strong2007, Vla, Orlando2013, Gulli}.
An exhaustive description of most recent improvements can be found in \cite{Moskalenko2015}.  
The GALPROP code computes CR propagation by numerically solving the
CR transport equation  over a grid in coordinates $(R, z, p)$, where $R$ is the radius from
the Galactic centre, $z$ is the height above the Galactic plane, and $p$ is the particle momentum.
The transport equation is described by the following formulation:

\newcommand{\drv}[2]{\frac{\partial #1}{\partial #2}   }
\newcommand{\opdrv}[1]{\frac{\partial}{\partial #1}  }
\newcommand{\ddrvm}[3]{\frac{\partial^{2}  #1}{\partial #2 \partial #3}  }
\newcommand{\ddrv}[2]{\frac{\partial^{2}  #1}{{\partial #2}^2}   }
\newcommand{\Dpp}{D_{pp}}
\newcommand{\Dxx}{D_{xx}}
\newcommand{\ddp}{{\partial\over\partial p}}

\begin{eqnarray}
{\partial \psi (\vec r,p,t) \over \partial t} 
&= &
  q(\vec r, p, t)                                             
   + \vec\nabla \cdot ( D_{xx}\vec\nabla\psi - \vec V\psi )   \nonumber  \\
&   +& \ddp\, p^2 D_{pp} \ddp\, {1\over p^2}\, \psi                  
   - {\partial\over\partial p} \left[\dot{p} \psi
   - {p\over 3} \, (\vec\nabla \cdot \vec V )\psi\right]  + \nonumber  \\
&   - & {1\over\tau_f}\psi - {1\over\tau_r}\psi\ 
\label{eq2}
\end{eqnarray}
where, the terms on the right side represent respectively: CR sources (primaries and secondaries), diffusion, convection (Galactic wind), diffusive reacceleration by CR scattering in the interstellar medium, momentum losses (due to ionization, Coulomb interactions, bremsstrahlung, inverse Compton and synchrotron processes), nuclear fragmentation and radiative decay. 
$\psi (\vec r,p,t)$ is the CR density per unit of total
particle momentum $p$ at position $\vec r$, $\psi(p)dp = 4\pi p^2 f(\vec p)dp $ in terms of
phase-space density $f(\vec p)$, $q(\vec r, p)$ is the source term including primary, spallation and decay contributions,
$\Dxx$ is the spatial diffusion coefficient and  is in general a function of  $(\vec r, \beta, p/Z)$ where $\beta=v/c$ and $Z$ is the charge, and $p/Z$ determines the gyro-radius in a given magnetic field. The secondary/primary nuclei ratio is sensitive to the
value of the diffusion coefficient and its energy dependence.
A larger diffusion coefficient leads to a lower ratio
because the primary nuclei escape faster from the Galaxy
producing less secondaries. Typical values of the
diffusion coefficient found from  fitting to CR  data are $\Dxx\sim(3 - 5)\times10^{28}$ cm$^{2}$ s$^{-1}$ at energy $\sim$1~GeV/nucleon increasing with magnetic rigidity as $\Dxx\sim R^{1/3}$ where the value of the exponent is typical for a Kolmogorov spectrum \citep{Strong2007}. $\vec V$ is the convection
velocity,  is a function of   $\vec r$ and depends on the nature of the Galactic wind.
Diffusive reacceleration is described as diffusion in momentum space
and is determined by the coefficient $\Dpp$\ related to  $\Dxx$ by $\Dpp\Dxx\propto p^2$. Moreover, $\dot{p}\equiv dp/dt$
is the momentum gain or loss rate. 
The term in $\vec\nabla \cdot \vec V$ represents  adiabatic momentum  gain or loss in the
 non-uniform flow of gas.
$\tau_f$ is the time scale for loss by
fragmentation, and  depends on the total spallation cross-section and the gas density  $n(\vec r)$ that can be based on surveys of atomic and molecular gas. 
$\tau_r$ is the time scale for  radioactive
decay \citep{Strong2007}.\\
GALPROP can be run both in 2D or 3D propagation scheme. 
The code calculates the propagation of the different species of CRs. 
Various parametrizations of CR source distributions \citep{Gullisources} as well as various models of the Galactic
magnetic field \citep{Orlando2013}, gas distributions \citep{diffuse2}, and the ISRF \citep{Porter2008} are included in GALPROP for computing the interstellar emission.
Even though numerical codes such as GALPROP contain many approximations, diffusion works well and allows hypotheses to be tested against different data.

\subsubsection{CR propagation models}
Our work aims at studying the following three {\em baseline propagation models} that we call PDDE, DRE, and DRC. For each
of these models we adopt the hadronic CR injection spectrum and the propagation parameters as described in greater
detail here below. 
Even though these are not the only possible propagation models, they represent the continuation of our previous works where propagation parameters for hadrons were inferred with dedicated fitting techniques and they were fitted to the latest {\em Voyager~I} data. Moreover they were made publicly accessible. 

\begin{enumerate}
\item {\em PDDE}:
We adopt the hadronic best-fit CRs injection spectra and propagation parameters from the very recent work by \cite{Voyager}. 
This corresponds to their plain diffusion model. The proton and helium injection spectra were fitted to
data from {\em Voyager~I} and PAMELA \citep{Adriani2011}. Heavier nuclei were fitted to {\em Voyager~I}, ACE-CRIS
\citep{ACE}, HEAO-3 \citep{HEAO}, and PAMELA \citep{Adriani2014},  
as described in detail in \cite{Voyager}. The tuning of the model parameters were performed in an iterative fashion
using the Minuit2 package from ROOT\footnote{http://root.cern.ch} by minimizing the $\chi^{2}$. Additional details on the fitting technique for the hadronic and isotopes are described in the appendix of \cite{Voyager}. 
A GALPROP plain diffusion model (and a diffusive-reacceleration 
model presented below as DRE model) with standard propagation parameters shows good agreement with {\em Voyager~1}
measurements of CR species from H to Ni in the
energy range 10 - 500 MeV/nucleon \citep{Voyager}. The reason of such an agreement may be the absence
of a recent source of low-energy CR hadrons in the solar
system neighborhood \citep{Voyager}. In the absence of such a CR
source, the shape of the spectra of CR species at low
energies is driven by the energy losses, mostly due to
the ionization, which are properly accounted for by the
GALPROP code. 
As discussed in the above paper among all-secondary Li, Be, and B nuclei, only B
measurements have a couple of low-energy data points
below 30 MeV/nucleon that show an excess over the
model predictions. 
Here the diffusion coefficient in the PDDE model must
decrease as the energy increases up to $\sim$4 GV in order to fit the
B/C measurements below 1 GeV nuc$^{-1}$. It is suggested \citep{Voyager} that a possible physical
justification of such behavior of the diffusion coefficient involves damping of interstellar turbulence due to the
interactions with low-energy CRs \citep{Ptuskin}. \\
We run GALPROP with these parameters, and the derived proton spectrum is shown in Fig.~\ref{fig1a}. Spectra of additional hadrons can be found in the original paper.

\begin{figure}
\includegraphics[width=0.4\textwidth]{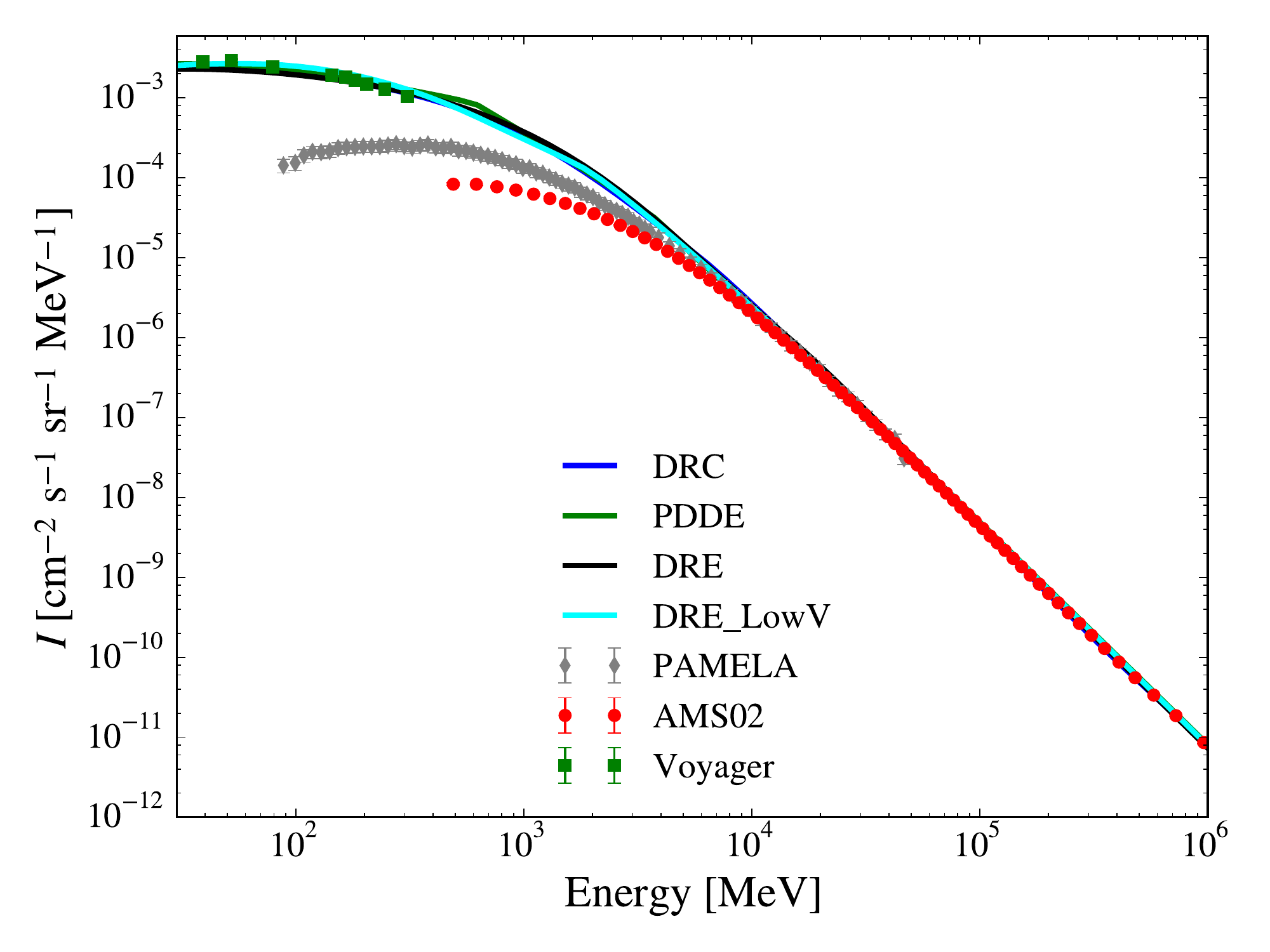}
\caption{Propagated proton LIS of the three baseline models DRE (black line), DRC (blue line), and PDDE (green line), plus DRELowV model (cyan line, described in Section 3.2.1) compared with data: red circles, AMS02 \protect\citep{AMSpro}; green squares, {\em Voyager~1} \protect\citep{Voyager}; grey diamonds, PAMELA \protect\citep{Pamelapro}. Propagation and hadronic injection parameters are as in \protect\cite{Voyager} for DRE and PDDE models, and as in \protect\cite{Boschini} for DRC model.}
\label{fig1a}
\end{figure}

\item {\em DRE}:
Also for this model we adopt the best-fit hadronic CRs injection spectra and propagation parameters from the very recent work by \cite{Voyager}.
This corresponds to the model with diffusion and reacceleration, which is statistically favored with high
significance with respect to the previous plain diffusion model ($PDDE$). More details on the modeling are described above and in \cite{Voyager}. We run GALPROP with these parameters, and the derived proton spectrum is shown in Fig. \ref{fig1a}. Spectra of additional hadrons can be found in the original paper.

\item {\em DRC}:
More recently, CR propagation models in the Galaxy were combined with propagation models in the heliosphere to reproduce direct measurements of CR hadrons at different modulation levels and at both polarities of the solar magnetic field \citep{Boschini}. A propagation model including diffusion, reacceleration, and convection was found \citep{Boschini} to give the best agreement with proton, helium, and antiproton data by AMS-02, BESS, PAMELA and {\em Voyager 1} from 1997 to 2015. The experimental observables included all published AMS-02 data on protons \citep{AMSpro}, helium \citep{AMSHelium}, B/C ratio 
\citep{AMSBtoC}. 
This is the most recent model for hadrons, where hadronic CRs and propagation parameters were fitted to AMS-02 and {\em Voyager~1} measurements. The HelMod\footnote{http://www.helmod.org/} code that computes the transport
of Galactic CRs through the heliosphere down to the
Earth was used. This provides a more physical treatment of the solar modulation instead of the force-field approximation. HelMod integrates the 
transport equation \citep{Parker} using a Monte Carlo approach that involves
stochastic differential equations. More details on HelMod are provided in \cite{HelMod} and \cite{HelMod2}, while on the joint implementation of HelMod with Galprop in \cite{Boschini}, where a MCMC procedure was used to
determine the propagation parameters. \\ 
The best-fit CRs injection and propagation parameters from that work are used to build our model with diffusion, reacceleration and convection. 
We run GALPROP with these parameters, and the derived proton spectrum is shown in Fig. \ref{fig1a}. Also here, spectra of additional hadrons can be found in the original paper.
\end{enumerate}

For the three models, PDDE, DRE, DRC, the propagation parameters are summarized in Table~\ref{Table1}. They are: $D_{0_{xx}}$, the normalization of the diffusion coefficient at the reference rigidity $D_R$; $D_{br}$, the rigidity break where the index of the rigidity can assume different values ($\delta_1$ and $\delta_2$); the Alfven velocity $v_{Alf}$; the convection velocity $v_c$, and its gradient $dV/dz$.

\begin{table}
\begin{center}
\caption{The table shows the propagation and the proton injection parameters of the models. Injection parameters for other nuclei are as in the original works \protect\citep{Voyager, Boschini} and are not repeated here. The description of each parameter can be found in the text.}
\begin{tabular}{lccccccc}
 \hline
 \hline
\\
         Model code & DRE  & DRC& PDDE &DRELowV$^{(b)}$\\                
  \\
 \hline
{Propagation} \\ {parameters }\\
\\
  D$_{0}$ $^{(a)}$ (cm$^2$~s$^{-1}$)& 14.6 & 4.3 &12.3     &  14.6   \\
  D$_{br}$ $(GV)$& -  & - &4.8  & -  \\
  $\delta_{1}$ & 0.327& 0.395 &-0.641  & 0.327  \\
  $\delta_{2}$ &0.323 & 0.395  & 0.578 & 0.323  \\
 V$_{Alf}$ (Km~s$^{-1}$)& 42.2& 28.6 &-  & 8.9 \\
 V$_{c}$ (Km~s$^{-1}$)& - & 12.4 & -& -   \\
 dV/dz (km~s$^{-1}$~kpc$^{-1}$) & - & 10.2 & -&-  \\
 \hline
{Proton} \\
{injection}\\
{parameters}\\
\\
$\gamma_{1}$  &0.65 & 1.69 & 1.18 & - \\
$\gamma_{2}$  & 1.94 & 2.44 & 2.95   & 1.4  \\
$\gamma_{3}$  & 2.47 & 2.28 & 2.22 & 2.47 \\
E$_{br_1}$ $(MV)$ & 117 & 700 & 124 & -  \\
E$_{br_2}$ $(GV)$ & 17.9 & 360.0 & 6.5 & 2.7 \\
\hline
\label{Table1}
\end{tabular}
\end{center}
$^{a}$ D$_{xx}$=10$^{28} \beta D_0(R/ D_R)^\delta$ cm$^2$ s$^{-1}$, with $D_R$=4GV for DRC model, and $D_R$=40GV for the other models. \\
The propagation halo size is 4 kpc for all the models. \\
$^{b}$ This propagation model is described in Section 3.2.1.\\
\end{table}

Model fitting of all-electrons was not addressed in the works by \cite{Voyager} and \cite{Boschini}. 
In this present work we infer injection spectral parameters of primary electrons to reproduce the CR all-electron measurements by {\em Voyager~1} and AMS-02, together with multifrequency data where possible. The resulting electron injection parameters will be given and discussed in Section 3.

\subsubsection{Interstellar emission calculations}

For each propagation model we generate the skymaps in the HEALpix scheme \citep{healpix} for the different interstellar emission mechanisms that are IC, pion decay, bremsstrahlung, and synchrotron. This is done by using the best 3D magnetic field formulation as found in \cite{Orlando2013} (as used in the so-called 'SUN10E' model in that paper), and the ISRF and gas model components as in \cite{diffuse2}. Regarding this latter component the conversion factor from CO to H$_{2}$ (X$_{C}$ $_{O}$) is assumed to be in the best-fit ranges as found in \cite{diffuse2} that better reproduces {\em Fermi} LAT gamma-ray data in the entire Galaxy. 
Specifically, for this conversion we make use of four Galactocentric rings having radii of 2, 6, 10, and 20 kpc with X$_{CO}$ values of 0.5, 6, 10, and 20 $\times$ 10$^{20}$ cm$^{-2}$(K km s$^{-1}$)$^{-1}$. 
An additional ingredient for computing the interstellar emissions is the distribution of CR sources, which is based on pulsars \citep{Lorimer} as in \cite{diffuse2}.  As suggested by {\em Fermi} LAT gamma-ray data \citep{outer2010, outer2011, diffuse2} and radio observations \citep{Orlando2013} we assume it to have a constant profile for a Galactocentric distance larger than 10 kpc. The IC emission is calculated with the anisotropic formulation of the Klein-Nishina cross section \citep{Moskalenko2000}.

\subsection{Observations}
For this study we use CR measurements and data from radio to gamma rays as described below. \\

\subsubsection{CR measurements}

Measurements of the CR spectra are affected by solar modulation below a few ten GeV only,
and until recently no CR data free from this effect were available below those energies. 
Since August 2012 {\em Voyager~1} observes a steady flux of Galactic CRs down to 3 MeV/nucleon for nuclei and to 2.7 MeV for all-electrons, which is independent on the solar activity. This is a strong indication of the instruments measuring the true LIS \citep{Voyager}.
We use {\em Voyager~1} all-electron measurements \citep{Voyager} together with the precise AMS-02 electron \citep{AMSele} and positron \citep{Accardo} data. 
PAMELA electron  
measurements \citep{Pamelaele} are also used for additional constraints. 

\subsubsection{Radio surveys}
Building upon the successful approach of \cite{Strong2011} we make use of those ground-based radio surveys at frequencies between 45 MHz and 1420 MHz, which display a nearly complete sky coverage ($>$80 per cent) in the region of interest. 
In the following we describe the single maps in more detail. At lowest frequencies the 45 MHz North map \citep{45MHz} and the  South map \citep{45MHzS} were combined to obtain an all-sky map by \cite{Guzman} with an offset of 500K. At somewhat higher frequencies, we adapt the 150 MHz map from the Parkes-Jodrell Bank all-sky
survey \citep{150MHz}. At 408 MHz the Haslam map \citep{haslam1981,haslam1982} as reprocessed by \cite{Remazeilles} is used in this work. We corrected this map by subtracting an offset of 8.9 K following the recent studies by \cite{Wehus2017}, \cite{CompPlanck} and \cite{LFPlanck}, which are found to be in agreement with our previous work \citep{Orlando2013}. The 408 MHz map is the only full-sky radio map with limited contamination from thermal emission. In addition, 
instrumental effects and sources have been accurately removed. These properties make this map an ideal tracer of the synchrotron radio emission from the Galaxy.
At higher frequencies the combined 1420 MHz North map from \cite{Reich82} and South map from \cite{Reich2001} are corrected for an offset of
3.28 K as computed in the very recent work by \cite{Wehus2017}. This value is in agreement with an
exhaustive work by \cite{Fornengo}. Offsets represent the sum of any instrumental and data processing offsets, as well as any Galactic or extra-Galactic components that are spectrally uniform over the full sky, including the CMB contribution. \\
To spectrally compare our propagation models with data we use the region of intermediate latitudes (i.e. 10$^\circ$$<|b|<$20$^\circ$)
because this includes mostly the local emission within $\sim$~1~kpc around the Sun and, hence, it encodes the CR LIS. 
Moreover, the region of intermediate latitudes is optimal because the synchrotron emission is the least contaminated: for $|b|<$20$^\circ$ offsets are not crucial even though we account for them, while for $|b|>$10$^\circ$ free-free absorption and emission are less than a few percent. However,
we remove this small contamination of the free-free emission by using the free-free spatial template released by the {\em Planck} Collaboration and by following the spectral formulation for the free-free emission as in \cite{CompPlanck}. We also account for the small contribution of the absorption using the implementation explained in detail in \cite{Orlando2013}.

\subsubsection{Microwave maps}
To study the synchrotron component we use the accurate four-year {\em Planck} synchrotron temperature map \citep{CompPlanck} released by the Planck Collaboration.
For an independent comparison we use also the nine-year Wilkinson Microwave Anisotropy Probe (WMAP) synchrotron maps \citep{Bennett}
at 23, 33, 41,
61 and 94 GHz obtained with the Maximum Entropy Method. 
While {\em Planck} provides the today's most accurate information on the synchrotron emission at microwave frequencies, the derivation of its intensity map is model dependent \citep{CompPlanck}. 
The derivation and relevance of {\em Planck} and {\em WMAP} maps will be discussed in Section 3.5. \\
Following the approach adopted for the radio surveys explained in Section 2.2.2, also the microwave synchrotron maps are used at intermediate latitudes (i.e. 10$^\circ$$<|b|<$20$^\circ$), excluding the Galactic plane where the contamination by free-free emission and anomalous microwave emission is important. In turn, this allows us comparing synchrotron spectra with models in a frequency range from a few tens of MHz to a few tens of GHz.

\subsubsection{Gamma rays}
The spectrum of the gamma-ray emission with its interstellar components (pion decay, bremsstrahlung and inverse Compton) encodes the spectra of CRs in the Galaxy.
A detailed study of the interstellar emission from the whole Galaxy was performed on a grid of 128 propagation models \citep{diffuse2} using the Fermi-LAT data. Even though all models provide a good agreement with data, no best model was found.
That study extensively investigated many GALPROP CR propagation models accounting for uncertainties in the models, such as ISRF, gas distribution, HI spin temperature, propagation halo size, and CR source distribution. However, it investigated propagation models with reacceleration only, which are challenged by synchrotron data \citep{Strong2011, Jaffe}. 
Here we test how different propagation models (i.e. DRE with reaccelereation, PDDE plain diffusion, and DRC with convection) spectrally compare with gamma-ray data. 
As a first step we use {\em Fermi} LAT gamma-ray spectra obtained in the study of \cite{diffuse2} for intermediate latitudes (i.e. 10$^\circ$$<|b|<$20$^\circ$).
For the purpose of comparisons, models are treated like data, i.e. integrated and averaged in the same
sky region. \\
In a second step, we use a specific dataset: the local HI gamma-ray emissivity. This directly encodes the spectra of CR LIS. The derivation of the emissivity
requires a careful approach. Such an approach has been followed in a recent work
\citep{JM}. In it the HI emissivity for the mid-latitude (10$^\circ$$<|b|<$70$^\circ$)
band, which is considered local, is derived by using {\em Fermi} LAT P7 reprocessed data
having energies between 50 MeV and 50 GeV that were taken in 4 years of observations, based on the extensive analysis in \cite{IEM}. 
This work \citep{JM}
carefully accounts for the {\em Fermi} LAT energy dispersion, which impacts the spectrum below
a few hundred MeV. It accounts also for large-scale structures such as the North Polar Spur
\citep{haslam1981}, the so-called Fermi bubbles \citep{Su, Dobler, FermiBubbles}, and the Earth's Limb emission. 
In the derivation of the local HI emissivity and its error bands three major sources of systematic errors are properly accounted for: the HI spin temperature, the
modeling of the IC, and the absolute determination of the {\em Fermi} LAT effective area \citep{JM}.
This recent derived local HI emissivity is used in our
model comparisons.\\
As the last step, we look at the Galactic centre region by using {\em Fermi} LAT spectra obtained with 
6.5 yeas of observations 
that were very recently
published in \cite{P8IG}. In this work, the original data are in flux units that we have converted
in intensity. For the purpose of comparisons, models are treated like data, i.e. integrated and averaged in the same
sky region, and masking out the most luminous sources as done to the original data.
These data are very suitable for qualitatively model comparisons of the 10$^{\circ}$ region around the Galactic centre. 
Due to the complexity in this region  we focus on the interstellar emission produced by the above propagation models neglecting the other components (i.e. isotropic, faint sources, solar and lunar, etc.), as reported in Section 3.4.

\subsubsection{X-rays and soft gamma rays}
At X-ray and soft gamma-ray energies data are taken by the INTErnational Gamma-Ray Astrophysics Laboratory 
({\em INTEGRAL}) mission \citep{winkler13} with its coded-mask telescope SPI, the SPectrometer for INTEGRAL\citep{SPI}.
In a detailed study by \cite{Bouchet}, spectral data of the Galactic diffuse emission are provided for energies between
$\sim$80 keV and $\sim$2 MeV. Data were taken for a very long integration time
ranging from year 2003 to 2009 for a total exposure of $\sim$10$^8$s on the sky
region $|b|<$15$^\circ$ and 
330$^\circ$$<l<$ 30$^\circ$. For the same sky region intensity data
at somewhat higher energies between 1--30 MeV are provided
by \cite{Strong1999} from the Imaging Compton Telescope (COMPTEL) instrument \citep{Comptel} on board of the Compton Gamma-Ray Observatory. 
Adopting the energy ranges from \cite{Strong1999}, maps are used
in three energy bands: 1--3 MeV, 3--10 MeV, and 10--30 MeV. \\
SPI and COMPTEL data were both cleaned by subtracting the sources \citep{Strong1999, Bouchet}.
For the limited sensitivity of those instruments at hard X-rays and MeV energy ranges, data in the inner Galaxy region, where the diffuse emission is maximum, are very suitable for model comparisons.

\section{Results}
This section presents results from the comparison of the GALPROP propagation models with CR
measurements and multi-wavelength data. 

\subsection{Baseline models}
For the three baseline models (DRE, DRC, PDDE) the propagation parameters for primary electrons are fixed to the values found for the hadronic propagation parameters.
The primary electron spectra parameters (injection spectral indexes and breaks) instead are inferred so that the all-electrons 
reproduce, after propagation, the precise data by AMS-02 above a few tens GeV and to reproduce the very recently
measured data by {\em Voyager 1} below 30 MeV. At the same time, primary electrons are inferred also to reproduce at best the synchrotron data (i.e.
radio and microwave surveys), as discussed in the next paragraph. PAMELA data data are used as an additional constraint: the LIS can not be lower than the direct measurements (being taken during solar minimum PAMELA measurements are higher than AMS-02 measurements). 
Positrons that contribute to the well-known 'positron excess' above 10~GeV are considered to originate from local sources\footnote{A further option to explain the positron excess is the dark matter scenario, which is investigated by many authors \citep[e.g.][]{Bertone}). For a recent review in the matter see \cite{Lipari}, while for a comprehensive review on CRs and their sources, see \cite{Funk, Blasi, Isabelle, Caprioli}.} \citep[e.g.][]{Mertsch, DiMauro, DellaTorre}. These sources are supposed to produce also the same amount of electrons. The contribution of these local electrons and positrons to the interstellar emission from radio to gamma energies is negligible. \\
Injection electron parameters are reported in Table~\ref{Table2}, with $\gamma_1$, $\gamma_2$, $\gamma_3$ spectral indexes, and $E_{br_1}$, $E_{br_2}$ energy breaks.

\begin{table}
\begin{center}
\caption{The table shows the electron injection parameters of the models. The description of each parameter can be found in the text.}
\begin{tabular}{lccccccc}
 \hline
 \hline
         Model code & DRE  & DRC& PDDE &DRELowV$^{a}$\\                
 \hline
$\gamma_{1}$  &2.90 & 2.75 & 2.01 & 2.20 \\
$\gamma_{2}$  & 0.80 & 0.65 & 2.55   & 1.70  \\
$\gamma_{3}$  & 2.65 & 2.62 & - & 2.68 \\
E$_{br_1}$ $(MV)$ & 320 & 400 & 65 & 170  \\
E$_{br_2}$ $(GV)$ & 6.3 & 4.0 & - & 4.5 \\
\hline
\label{Table2}
\end{tabular}
\end{center}
$^{a}$ This propagation model is described in Section 3.2.1.\\
\end{table}

\begin{figure}
\catcode`\_=12
\includegraphics[width=0.45\textwidth]{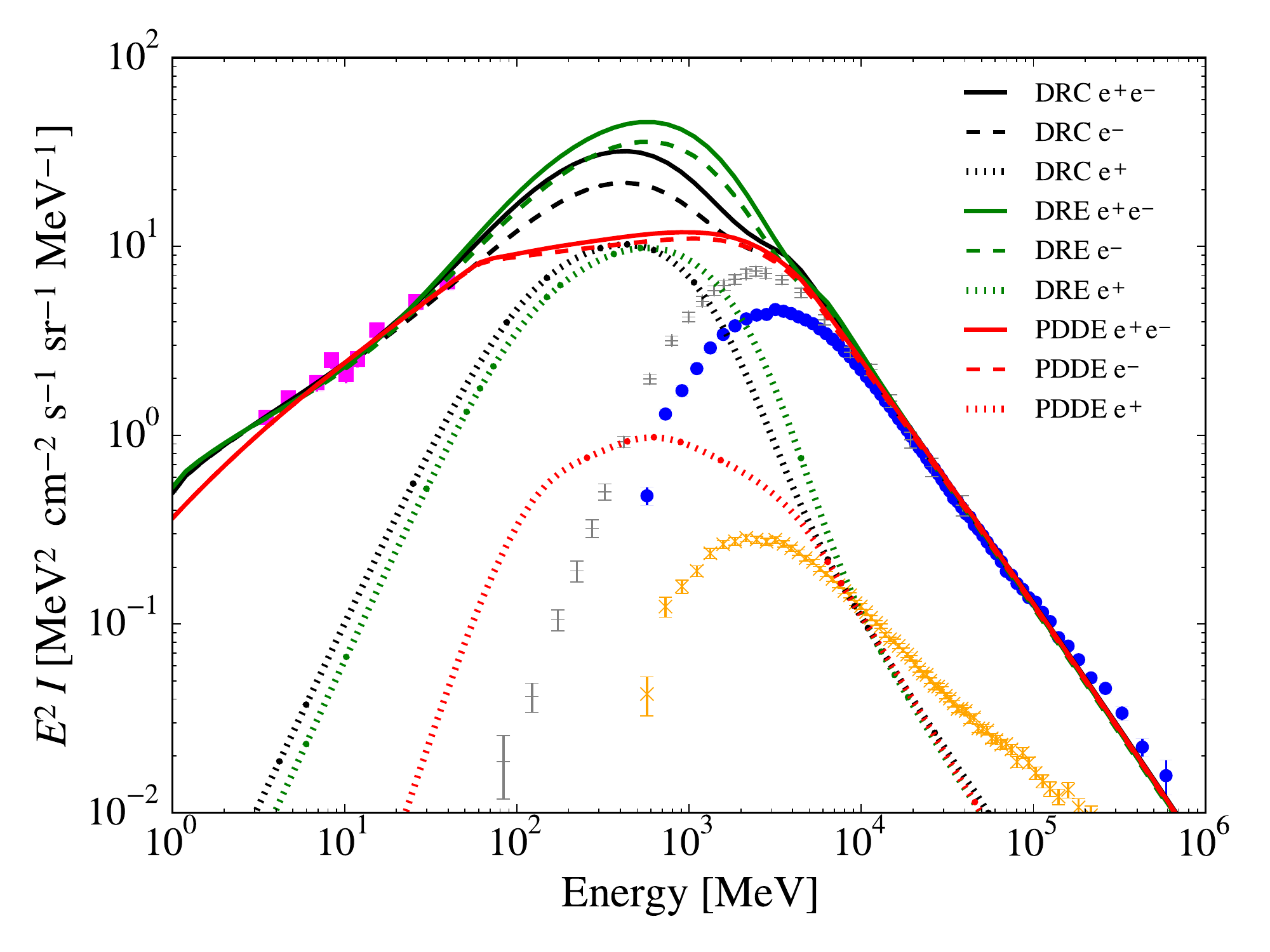}
\caption{Propagated interstellar spectra of the three baseline models DRE (green line), DRC (black line), and PDDE (red line) for positrons (dotted lines), electrons only (dashed lines), and  all-electrons (solid lines) compared with data: orange crosses: AMS-02 positrons \protect\citep{AMSele}; blue points: AMS-02 electrons \protect\citep{AMSele}; grey dashes: PAMELA electrons \protect\citep{Pamelaele}; magenta squares: {\em Voyager 1} all-electrons \protect\citep{Voyager}.}
\label{fig1}
\end{figure}

For the three propagation models (DRE, DRC, PDDE) Figure~\ref{fig1}  
shows the comparison of the propagated all-electron
LIS (solid lines), along with their distinct components of electrons (dashed lines) and secondary positrons (dotted
lines), with the direct CR measurements (squares for {\em Voyager 1}, dots for AMS-02 electrons, crosses for AMS-02 positrons, dashes for PAMELA electrons). 
The three baseline models produce three
different all-electron LIS densities in the range $\sim$(10$^{2}$--10$^{4}$)~MeV. In this range the low all-electron
density of the PDDE model (red line in Figure~\ref{fig1}) is due to the break of the diffusion coefficient, while the injection spectrum is the same
downwards to a few tens of MeV. Only below a few tens of MeV a break in the injection spectrum is necessary to
avoid overestimating {\em Voyager 1} data. On the other hand, the DRC (black line in Figure~\ref{fig1}) and the DRE (green line in Figure~\ref{fig1}) models require two
breaks in the primary electron injection spectra to reproduce {\em Voyager 1} data.
We can summarize by saying that models without breaks in the injection spectrum of primary electrons at low energies can not reproduce the {\em Voyager~1} data. 
It is worth noting that the contribution of secondary positrons in the range
$\sim$(10$^{2}$--10$^{4}$)~MeV for the models encoding reacceleration (i.e. DRC, DRE) is a factor of ten larger compared to the PDDE model. While the very similar proton spectrum among the three models can not account for this difference, reacceleration processes can.  \\

\begin{figure}
\catcode`\_=12
\includegraphics[width=0.45\textwidth]{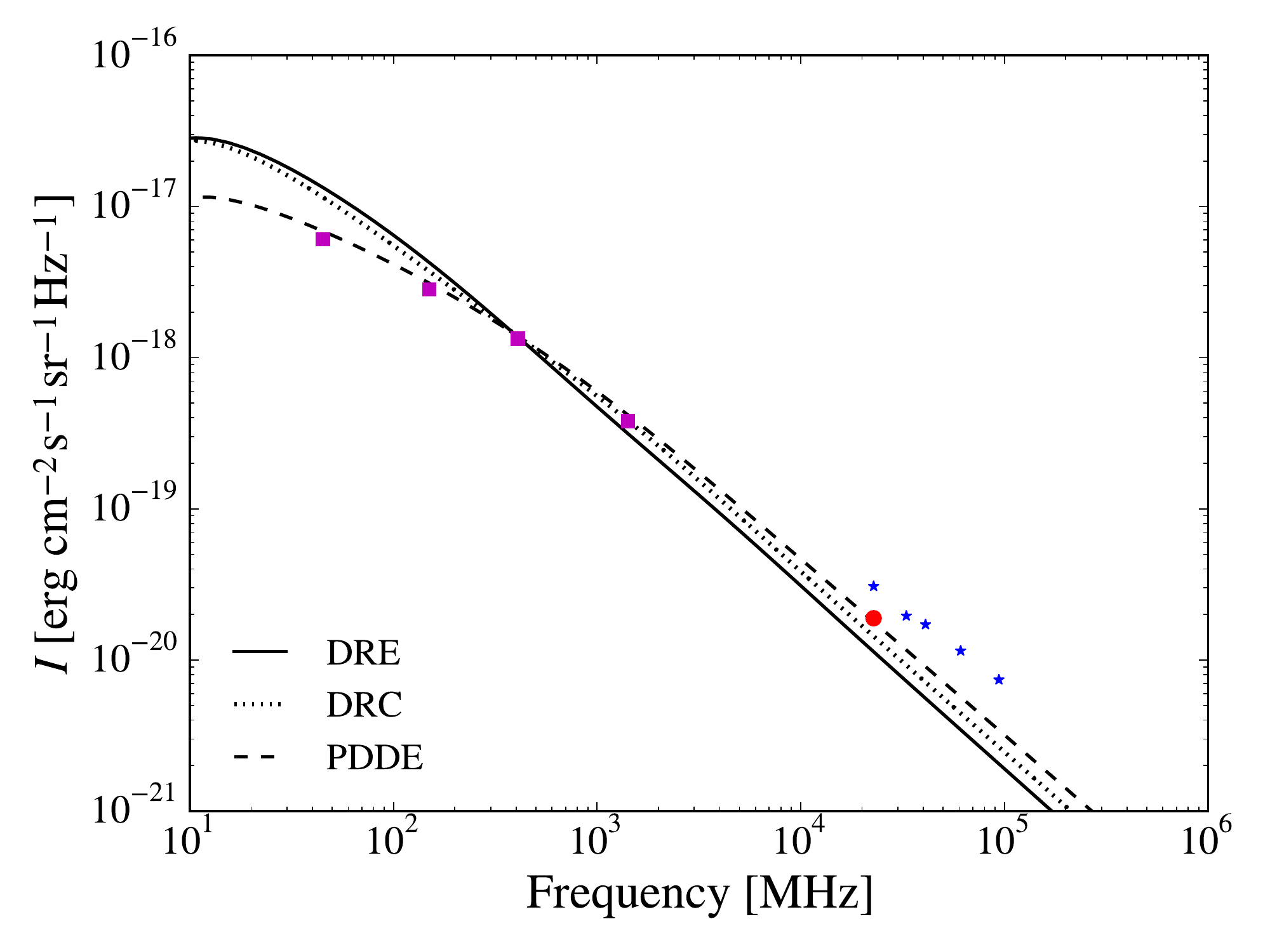}
\caption{Synchrotron spectra for intermediate latitudes (10$^\circ$$<|$b$|<$20$^\circ$) of the three baseline models DRE (solid lines), DRC (dotted line), and PDDE (dashed line) compared with data: radio surveys (magenta squares) (described in Section 3.2), {\em Planck} synchrotron map (red point) \protect\citep{CompPlanck}, and WMAP (blue stars) \protect\citep{Bennett}.  }
\label{fig2}
\end{figure}

We report here on the comparison of the calculated synchrotron spectra to the synchrotron data. As previously stated the primary electrons were tuned so that the all-electrons reproduce at best not only the direct CR measurements but also the synchrotron data for the energy range where the CR direct measurements are affected by solar modulation (i.e. $\sim$(10$^{2}$--10$^{4}$)~MeV).
To constrain the CR all-electrons with synchrotron data, we use the best-fit normalization of the magnetic field intensity found
by spatially fitting the calculated synchrotron template to the observed 408~MHz map, after subtracting
the free-free emission component and the offsets, as successfully performed in our previous work \citep{Orlando2013}. 
In Figure~\ref{fig2} we display the resulting synchrotron spectra of the three baseline
models (DRE solid line, DRC dotted line, PDDE dashed line) using the all-electrons as in Figure~\ref{fig1}. 
We compare the calculated spectra to the synchrotron emission by radio surveys and by the {\em Planck} synchrotron map integrated at intermediate latitudes.
While {\em Planck} provides the today's most accurate information on the synchrotron emission at microwave frequencies, {\em WMAP} maps are used as upper limits (see discussion on {\em Planck} and {\em WMAP} uncertainties in Section 3.5). 
Figure~\ref{fig2} shows that the synchrotron spectrum of the PDDE model performs best in the entire frequency range
compared to the DRE and DRC models that overestimate the observations at frequencies below 408~MHz.
The overestimation is due to the larger density of the all-electron LIS at $\sim$(10$^{2}$--10$^{4}$) MeV. 
This enhancement is due to strong reacceleration processes (with Alfven velocity around 30 -- 40 km/s) responsible to contribute to secondary CRs. 
This is in agreement
with our previous findings \citep{Strong2011}.
The same significant amount of secondaries prevents from tuning the primary electron spectrum of the DRE and DRC
models in such a way 
to reproduce the synchrotron intensity at 
$\sim$(10 -- 400)~MHz. At these frequencies the eye-catching gap between the DRE/DRC models and the PDDE model can be
seen in Figure~\ref{fig2}. To further investigate this difference we make use of the following additional approach. To avoid assumptions on primary electrons, we derive these by subtracting the secondaries, calculated with GALPROP for the three baseline models (DRE, DRC, PDDE), from the all-electron LIS that fits both synchrotron observations and CR measurements.
After the subtraction we are left with the
spectrum of primary electrons only, which can be compared to the electron direct measurements by PAMELA and AMS-02. As a result, the primary electron spectrum obtained for DRE and DRC models below a few GeV are either negative or null. This means that the spectrum of secondaries for the DRE and DRC model is larger or equal to the all-electron LIS that reproduces the synchrotron data.
This leaves no space for a meaningful primary electron spectrum of the DRE and DRC models. Instead,
for the PDDE model, the derived spectrum of primary electrons is in agreement with CR measurements. We can conclude that the
two independent approaches (i.e. the latter approach without assumptions on the primary electron spectrum, and the previous approach with
the tuning of it) lead to the same result: propagation models that produce significant amount of secondaries or that
have a large all-electron intensity in the range $\sim$(10$^{2}$--10$^{4}$)~MeV are difficult to reconcile with synchrotron data.\\ 
The values of the spectral intensity of all-electron LIS for our best model PDDE is reported in Appendix (Table~\ref{TableApp1}). \\

\begin{figure*}
\includegraphics[width=0.32\textwidth]{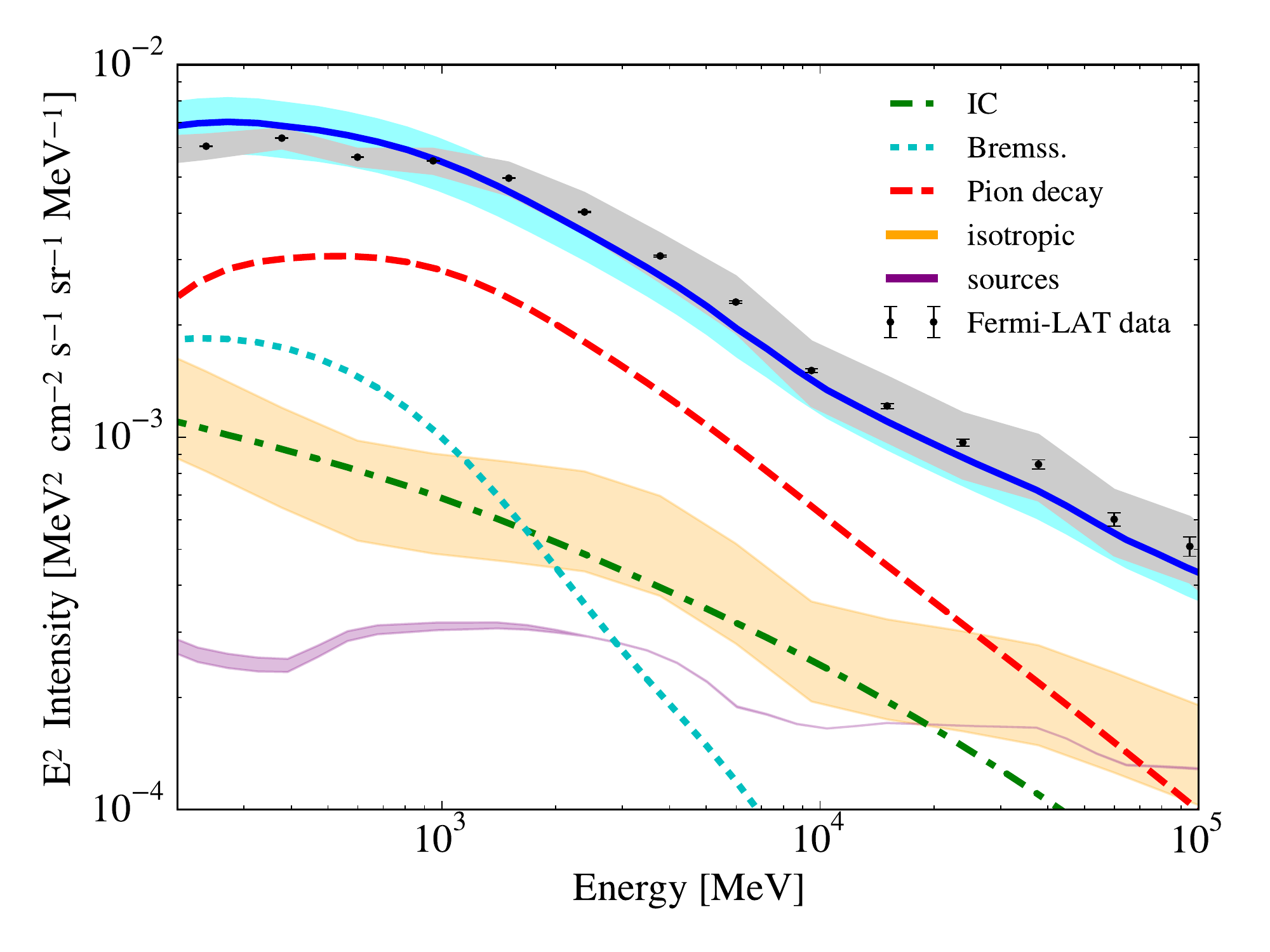}
\includegraphics[width=0.32\textwidth]{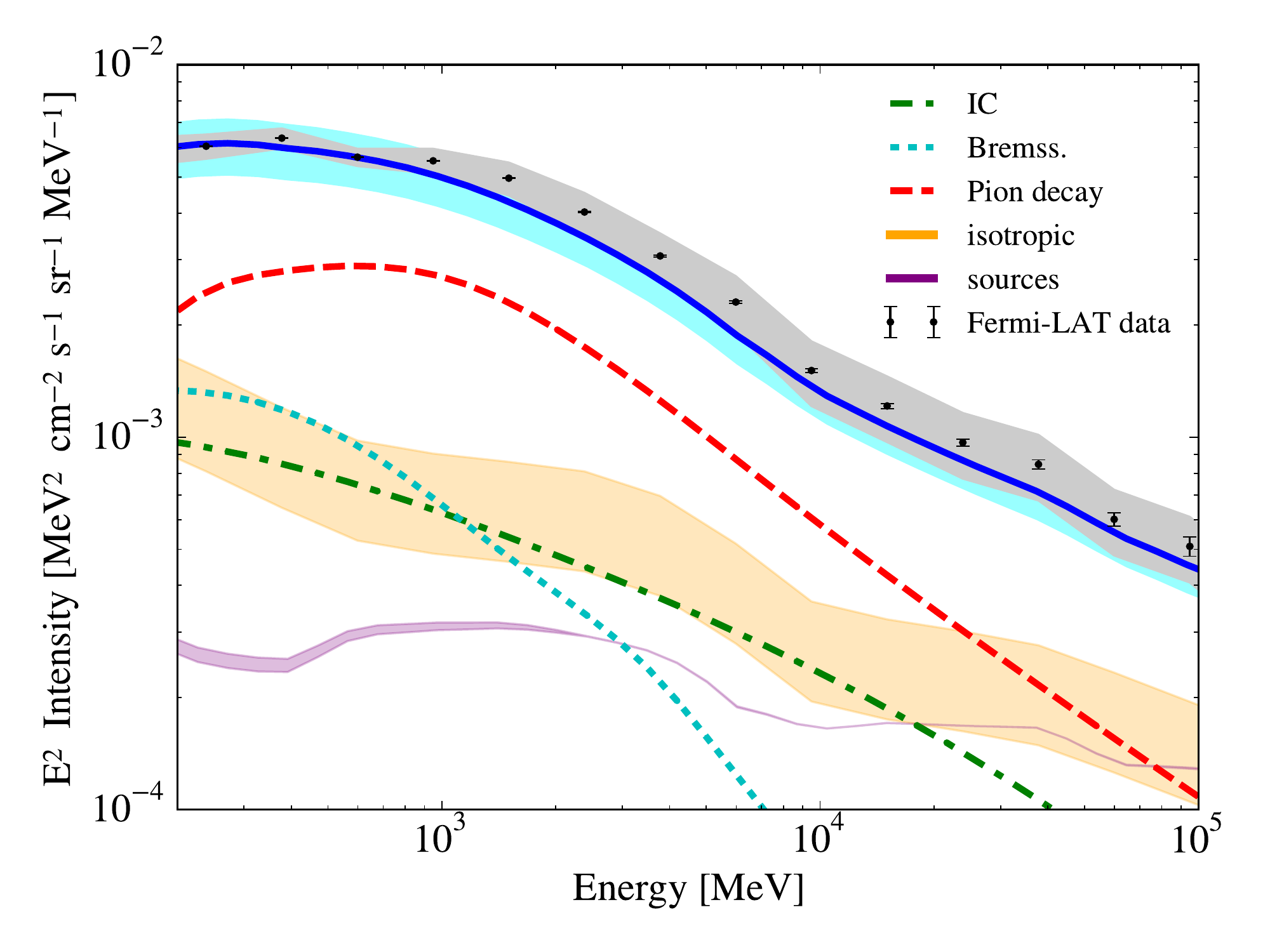}
\includegraphics[width=0.32\textwidth]{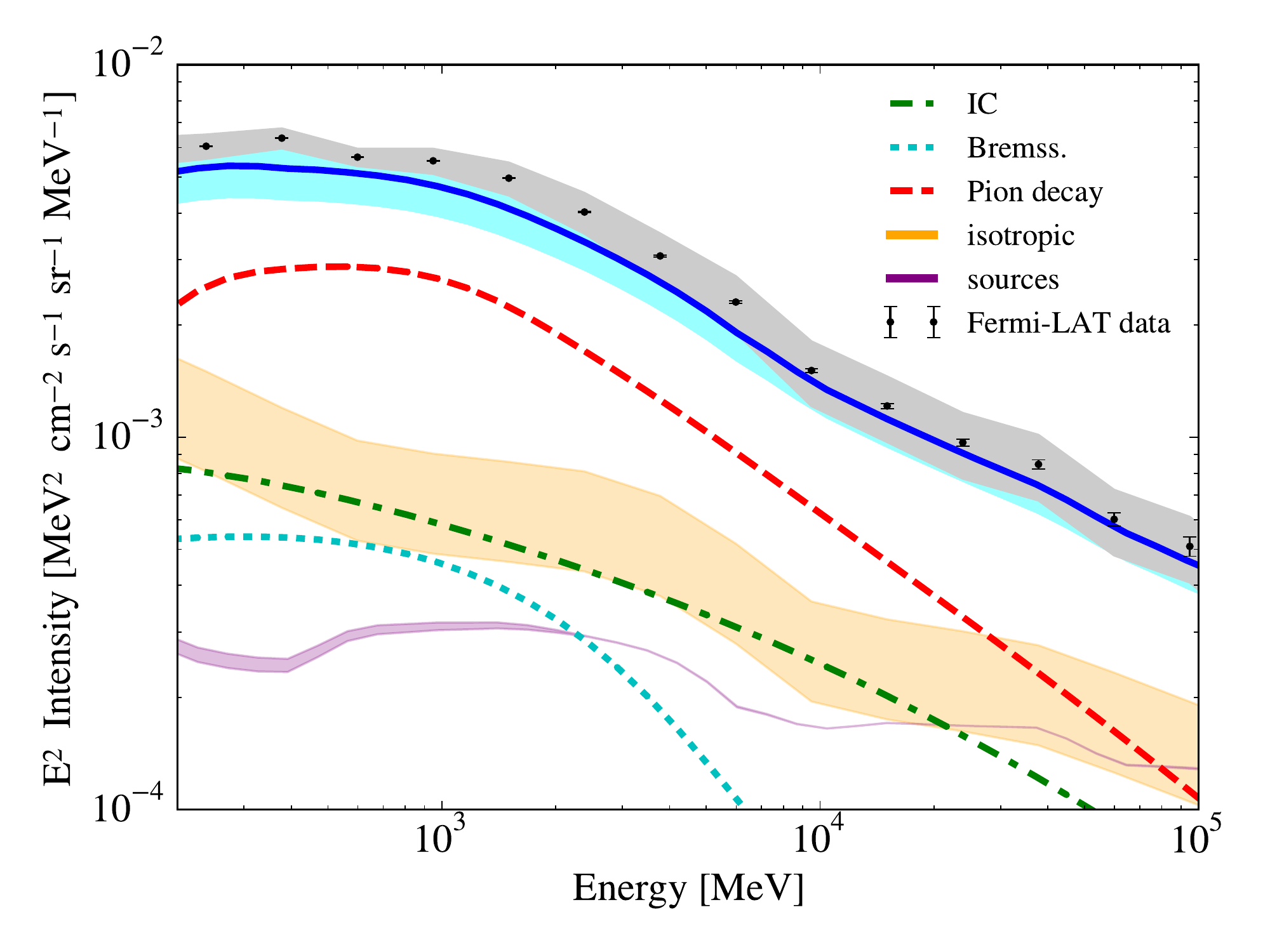}
\caption{Calculated gamma-ray spectral intensity of the three baseline models, left to right, DRE, DRC, and PDDE. Data are {\em Fermi} LAT spectra at intermediate regions (10$^{\circ}$ $<$|b|$< $20$^{\circ}$, all longitudes) from \protect\cite{diffuse2}. Models are: pion decay (red dashed line), IC (green dash-dotted line), bremsstrahlung (cyan dotted line). Data include statistical errors (grey area) and systematic errors (black bars). Spectra for sources (magenta region) and isotropic component (yellow region) are taken from \protect\cite{diffuse2}, for the most extreme cases reported there. Uncertainty of 30 per cent is added to the isotropic spectrum, following the study in \protect\cite{EGB} based on various foreground models. Interstellar model components are not fitted to data.}
\label{fig3a}
\end{figure*}

Gamma-ray data provide an additional source of information on the all-electron and proton spectrum. While in our previous work \citep{diffuse2} only reacceleration models (similar to our DRE model) were studied, here we spectrally test the different propagation models (DRE, DRC, PDDE) with gamma-ray data.
Hence, we calculate the gamma-ray emission expected from the three propagation models (DRE, DRC, PDDE) at intermediate latitudes. Figure~\ref{fig3a} shows the comparison of these predictions with {\em Fermi} LAT data for the intermediate latitudes as published in \cite{diffuse2}. Spectra for detected gamma-ray sources and for the isotropic emission are taken from \cite{diffuse2}, for the most extreme cases reported there. An uncertainty of 30 per cent is added to the isotropic spectrum, following the study in \cite{EGB} based on various foreground models. Below a few hundred MeV, DRE and DRC models produce higher gamma-ray emission than PDDE model due to the enhanced all-electron density, which in turn increases the bremsstrahlung emission. However, all the models are within the {\em Fermi} LAT systematic uncertainties. 
Hence, in a first approximation, with the data used here, also plain diffusion models, such as our PDDE model, reproduce gamma-ray data as well as reacceleration models. However, in general, analyses of the gamma-ray data in various regions of the Galaxy suffer from large uncertainties mainly given by the ISRF and the gas density \citep[e.g.][]{diffuse2, EGB, P7IG, IEM}. \\  
\begin{figure*}
\catcode`\_=12
\includegraphics[width=0.4\textwidth]{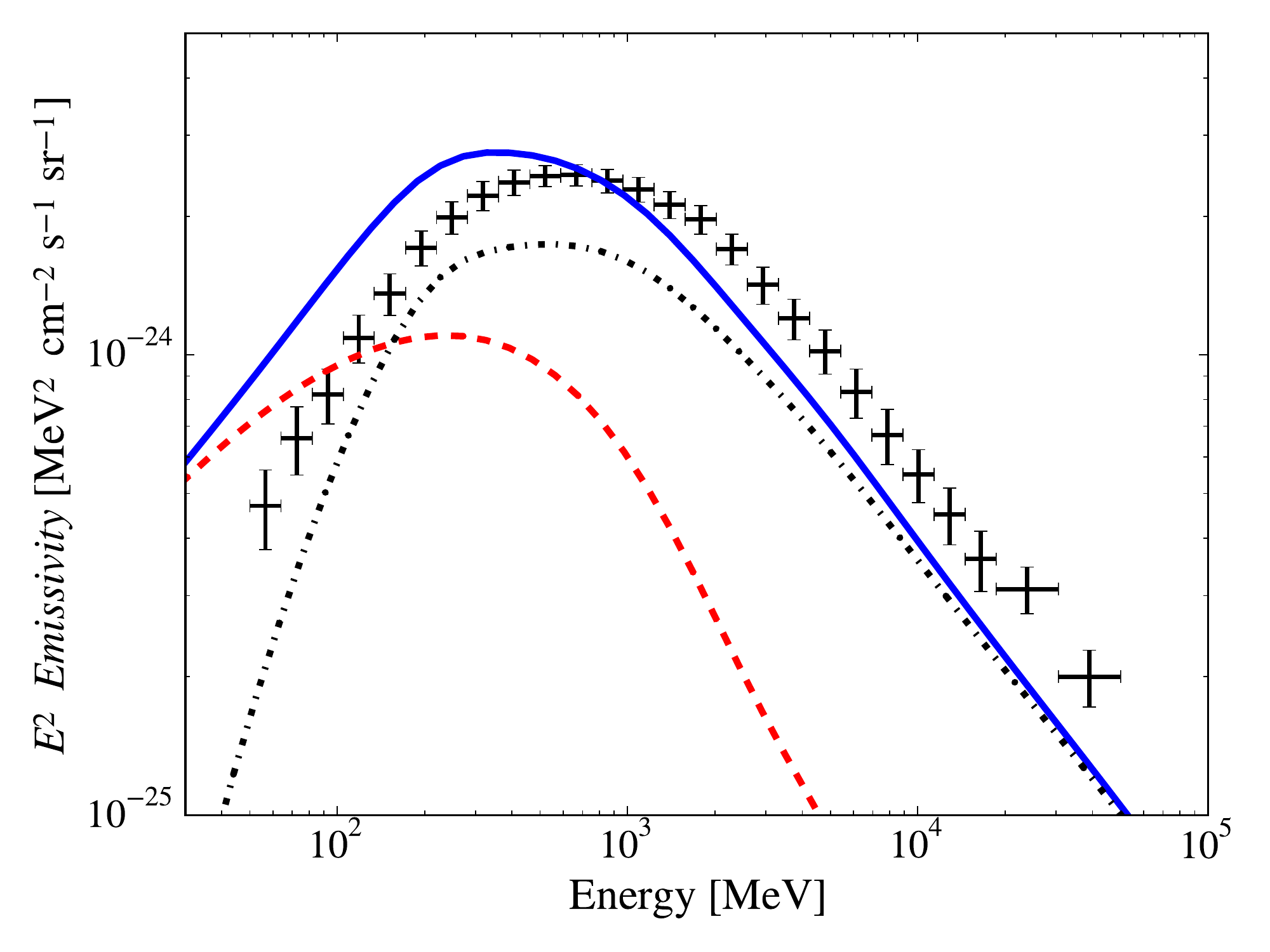}
\includegraphics[width=0.4\textwidth]{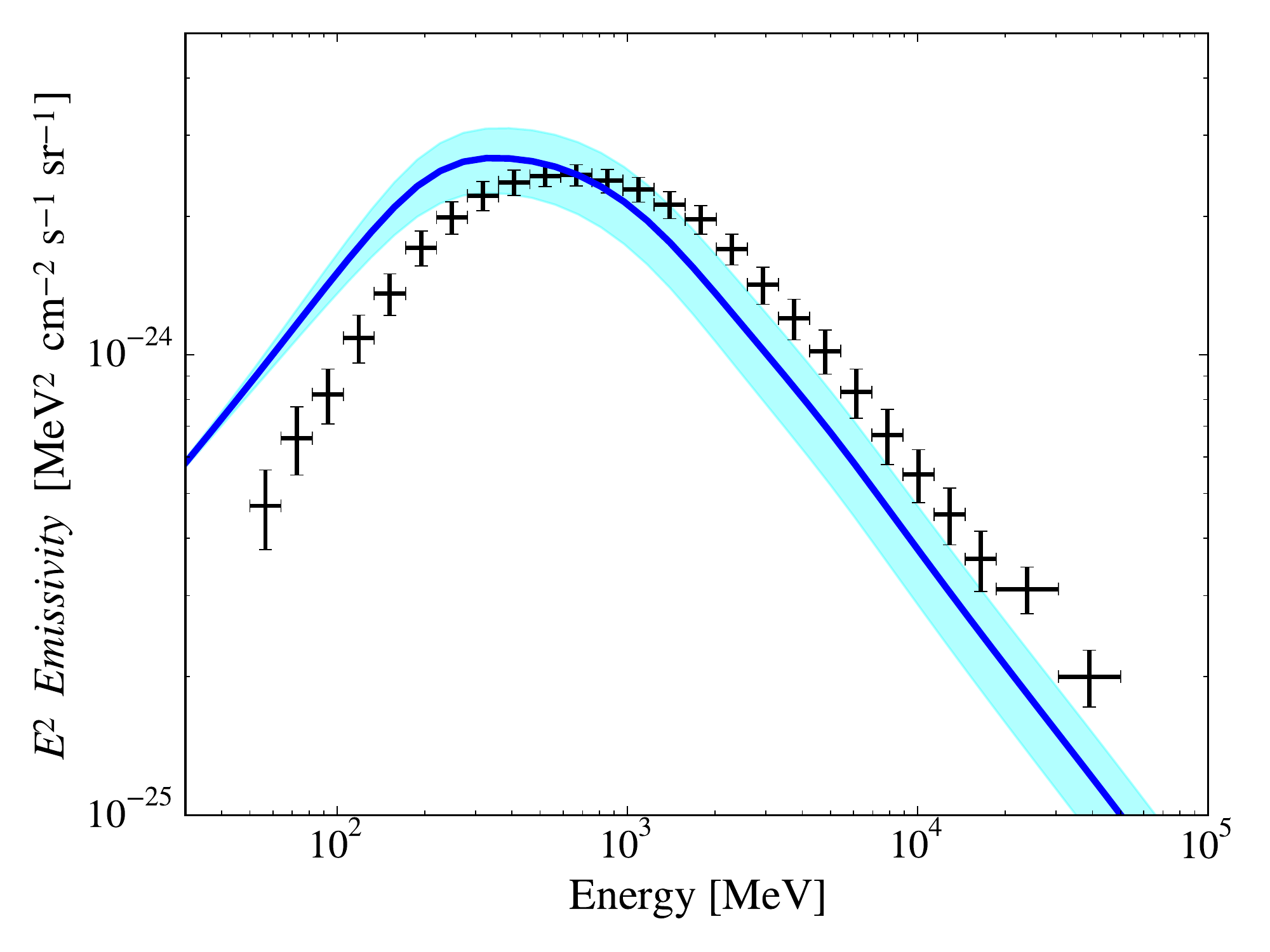}\\
\includegraphics[width=0.4\textwidth]{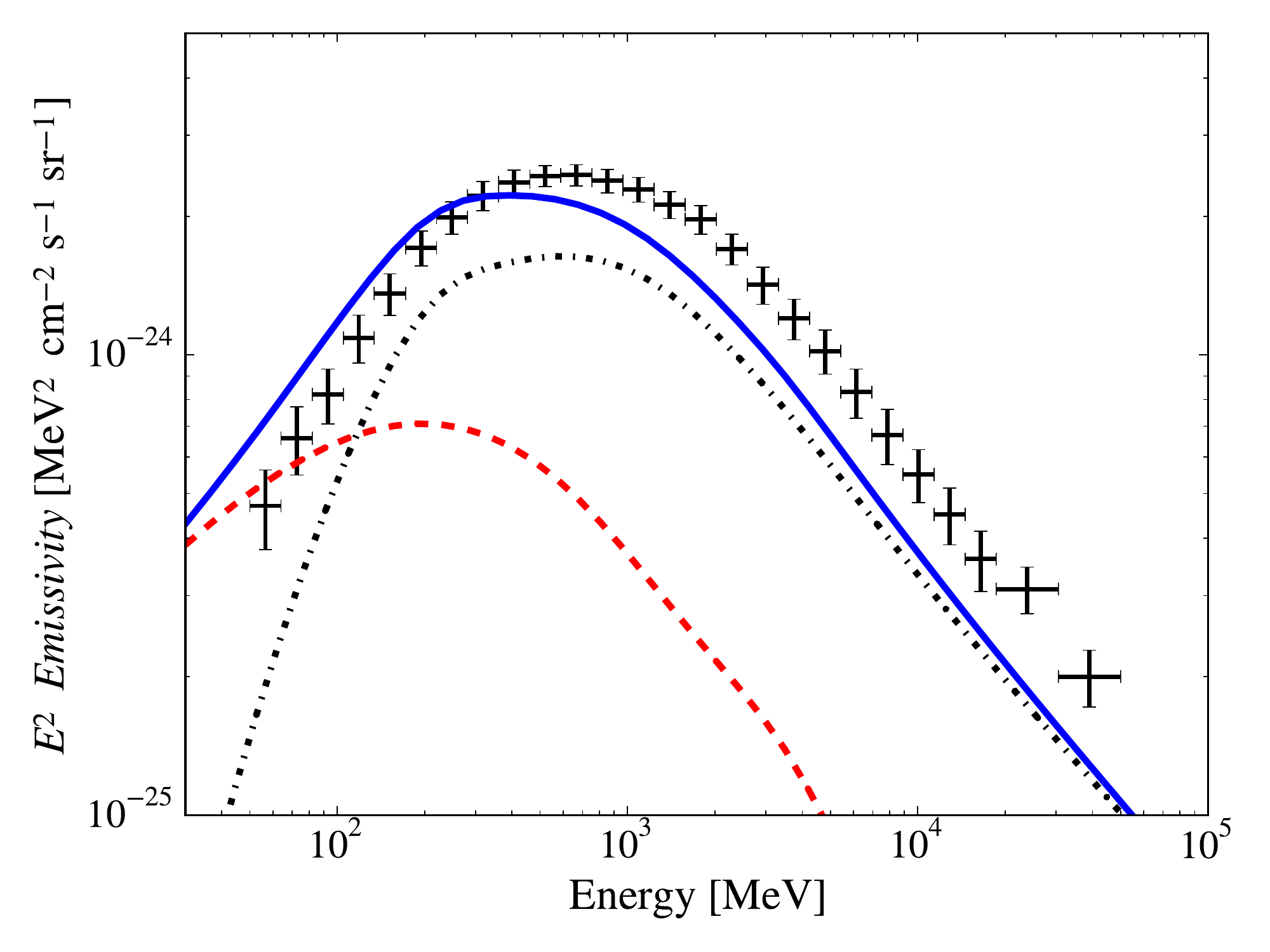}
\includegraphics[width=0.4\textwidth]{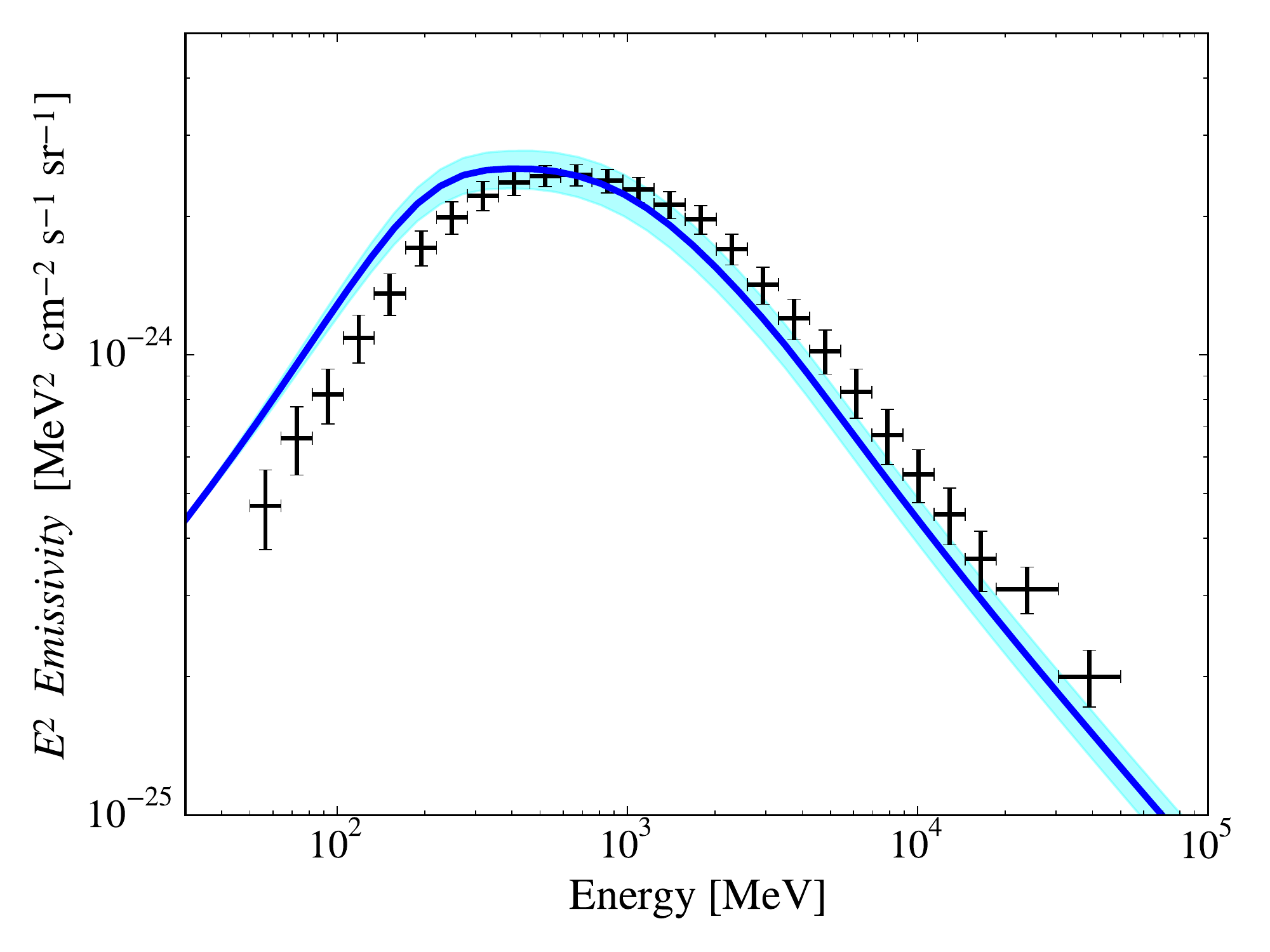}\\
\includegraphics[width=0.4\textwidth]{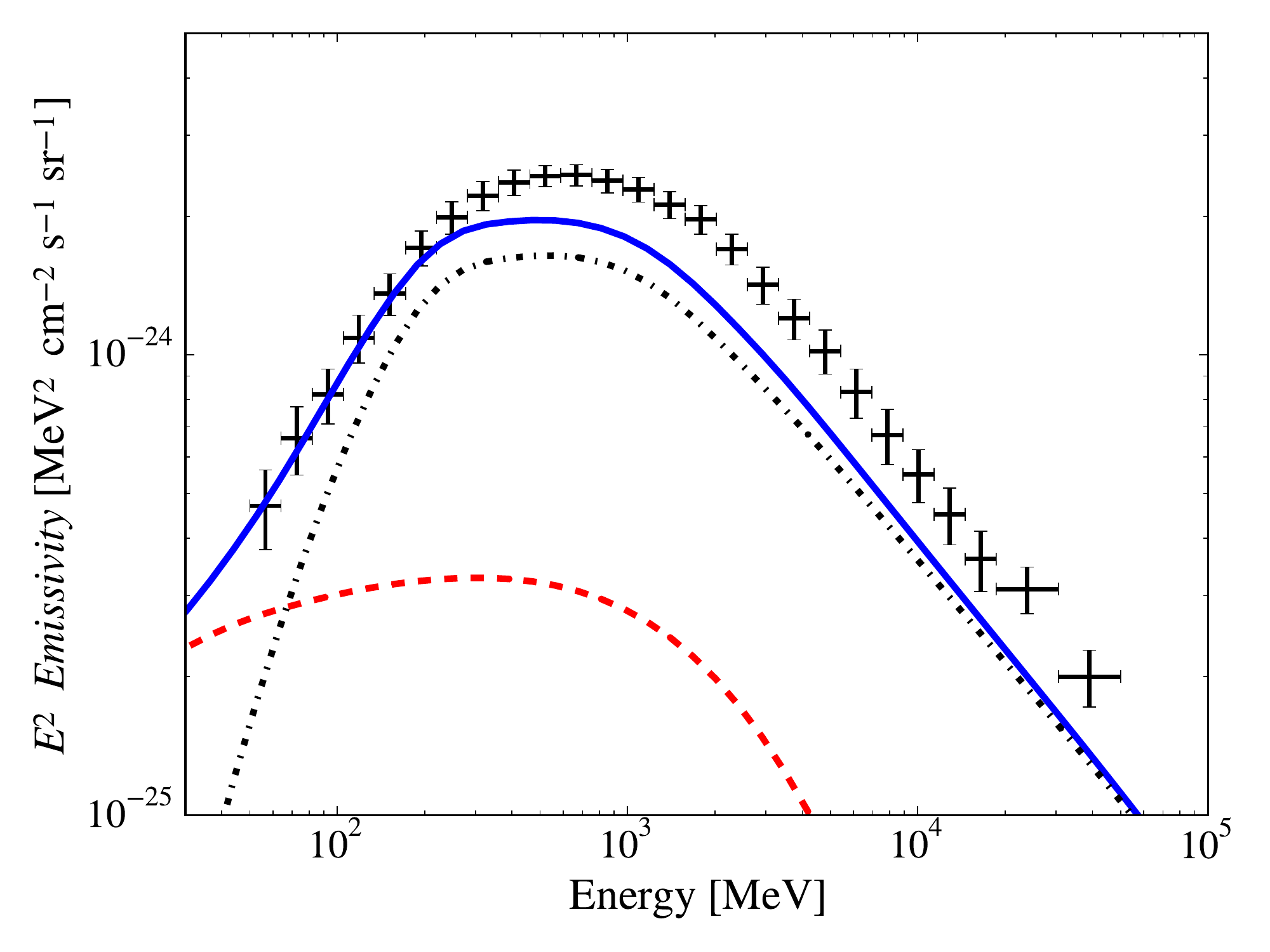}
\includegraphics[width=0.4\textwidth]{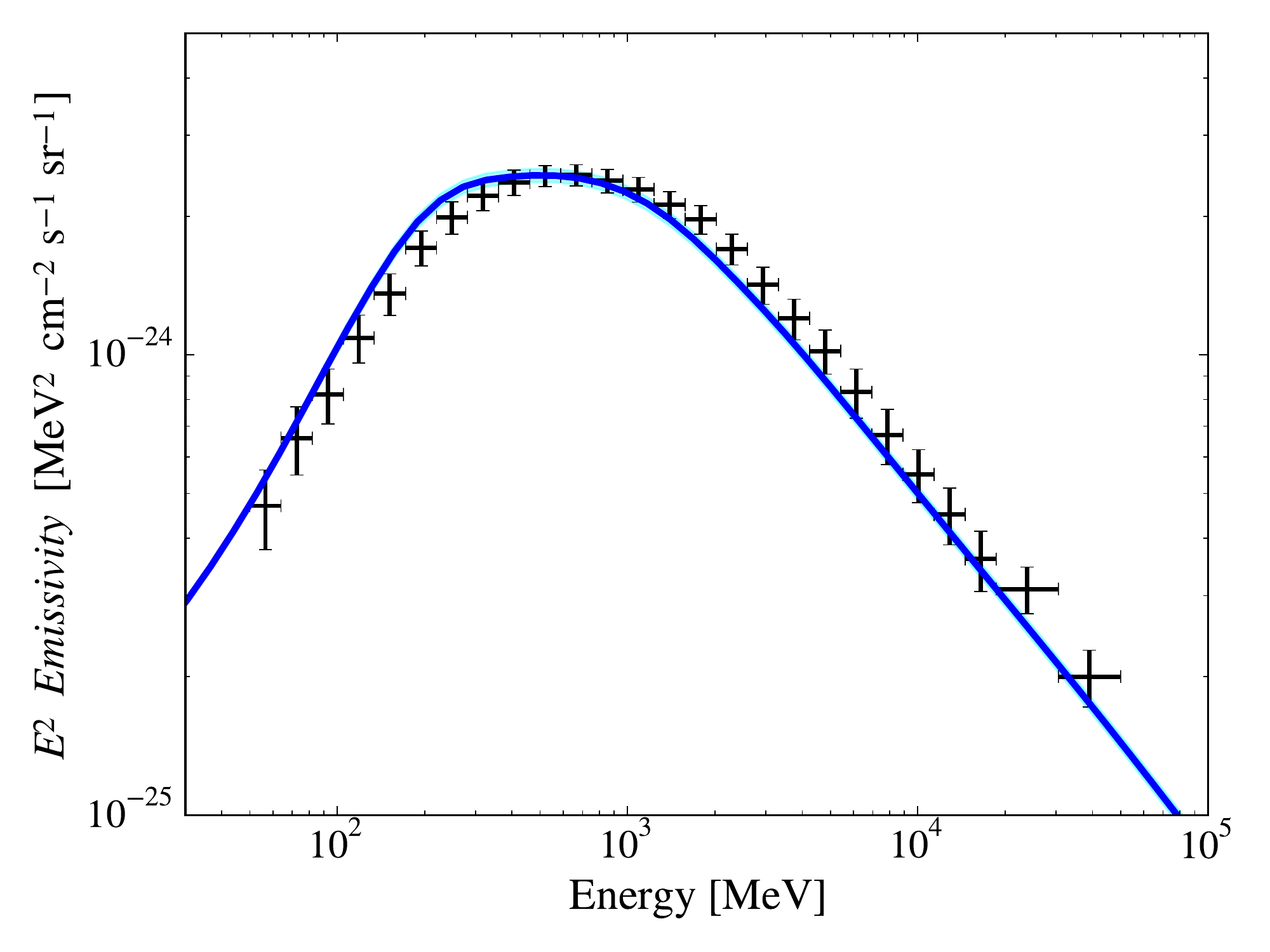}
\caption{Local gamma-ray emissivity of the three baseline models, top to bottom, DRE, DRC, and PDDE compared with {\em Fermi} LAT local HI emissivity \protect\citep{JM}. 
{\em Left}: Calculated bremsstrahlung components (red dashed lines), pion-decay components (blue dashed-dotted lines), and their sum (blue solid lines) are shown. {\em Right}: the sum (blue solid lines) of the calculated bremsstrahlung and the fitted pion-decay components, and the 1-$\sigma$ error (cyan region) in the fitted parameter are shown. Fitted parameters for the pion decay components, errors, and chi-squares of the fit are reported in Table~\ref{Table3} (normalization for the entire energy band).}
\label{fig3}
\end{figure*}
\begin{table}
\begin{center}
\caption{The table shows the best-fit values of the pion decay component to the gamma-ray emissivity.}
\begin{tabular}{lccc}
\\
 \hline
 \hline
 \\
 {Model} &   {Normalization }         & {chi-square}      &  {Normalization} \\
 &   {(entire energy band)}         &       &  {(> 1 GeV)}
  \\
 \hline
\\
DRE & 0.95 $\pm$ 0.26 &  10.5 & 1.38 $\pm$ 0.07\\
DRC & 1.20 $\pm$ 0.15& 4.2 & 1.45 $\pm$ 0.03\\
PDDE &  1.30 $\pm$ 0.05& 1.2 & 1.40 $\pm$ 0.05\\
DRELowV$^{a}$ & 1.35 $\pm$ 0.05& 0.6 & 1.40 $\pm$ 0.05 \\
\hline
\label{Table3}
\end{tabular}
\end{center}
$^{a}$ This propagation model is described in Section 3.2.1.\\
\end{table}
A precise way to obtain information about CR spectra and density in various places in the Galaxy is to study the emissivity per H atom that reflects the CR spectra free from uncertainties on the ISRF and gas distributions \cite[e.g][]{outer2010, outer2011,Tibaldo}. The HI emissivity includes the bremsstrahlung and pion decay components. A recent study was performed by \cite{JM}, which derived the local HI emissivity.
We examine our baseline models by comparing the calculated gamma-ray emissivity at the location of  $\sim$1~kpc around the sun
to the local HI emissivity data from {\em Fermi} LAT \citep{JM}. 
To facilitate the comparison between data and models, in Figure~\ref{fig3} we display from top to bottom the three baseline models
DRE, DRC, and PDDE. Plots on the left show the calculated components compared to the data, while plots on the right show the result of fitting the components to the data. In all of the three plots on the left, {\em Fermi} LAT data \citep{JM} are black crosses, the sum of the calculated components are solid lines,
their bremsstrahlung component is dotted, and their pion-decay component is dashed. For the first two plots on the left is it clear that the DRE and DRC models (blue solid lines) overestimate the data below several hundred MeV (black crosses). More strikingly, even their bremsstrahlung
component (dotted line) alone overestimates the data below 100~MeV. This finding strongly disfavors the DRE and the DRC models. Instead
the PDDE model (bottom plot in Figure~\ref{fig3}) reproduces the data very well below a few hundreds of MeV. 
This finding, that models with relatively low all-electron intensity below a few GeV reproduce gamma-ray data, reinforces the previous results where the same low all-electron 
intensity reproduces the radio observations. 
Above a few hundreds of MeV, for all the models the predictions of the local emissivity fail to reproduce
the {\em Fermi} LAT observations (left plots in the same figure).
To quantify this difference between data and models, Table~\ref{Table3} reports the best-fit scaling factors of the 
pion decay
components for the three models (DRE, DRC, PDDE, plus one model discussed later). The fit is performed by freezing the normalization of the bremsstrahlung component and leaving the normalization of the pion decay component free to vary. The chi-square values reported in Table~\ref{Table3} are significantly better for the PDDE model over the DRE and DRC models, which poorly fit the data (see also the right plots on Figure~\ref{fig3} especially below a few hundred MeV). DRE and DRC models still overestimate the data below a few hundred MeV, thereby being disfavored by data. Regarding the PDDE model, to match the measured data the numerical value suggests that the pion decay requires an increase of at least 30 per cent. 
Table~\ref{Table3} reports also the best-fit scaling factors of the pion decay component performed above 1~GeV only, where the contribution of the bremsstrahlung component is not significant. 
The best-fit scaling factors are around 1.3 -- 1.4 for all the models. 
Beside preferring the PDDE model, this comparison of the calculated emissivity with the observed emissivity suggests that the direct CR measurements do not represent the average spectrum in the local region within $\sim$~1~kpc  probed  by the observed local gamma-ray emissivity, even if solar modulation is taken into account.
Hence, we derive the proton spectrum that best reproduces the emissivity, based on the best-fit value reported in Table~\ref{Table3} for PDDE model (for the entire energy range). Our resulting proton LIS (red solid line) is compared to AMS-02 proton measurements (black points) in Figure~\ref{fig4}. The red region includes 10 per cent uncertainty on the cross sections \citep{JM} and the uncertainty in the fit parameter estimation (Table~\ref{Table3}). The discrepancy between our LIS based on the emissivity and the CR measurements from AMS-02 is evident even beyond the influence of the solar modulation.
Above a few GeV our normalized proton spectrum is general agreement with a recent work by \cite{emissivity2} on behalf of the Fermi LAT collaboration and with an earlier work by \cite{emissivity}, in which the proton LIS has been obtained from the local gamma-ray emissivity in a
model independent approach. Their complementary approach independently supports our results. However, the discrepancy data-model in \cite{emissivity2} was not found to be as strongly significant as we instead find now because in that work the proton spectrum derived from the emissivity was compared to PAMELA data, which have larger uncertainties than AMS-02 (more than 20 per cent uncertainties in PAMELA data with respect to 5 per cent uncertainty maximum in AMS-02). 
This most likely prevented \cite{emissivity2} from drawing definitive conclusions upon. 
The same figure also shows the best-fit LIS from \cite{emissivity2} and \cite{JM} for comparison (uncertainties are not plotted), supporting our conclusion that latest precise CR proton measurements do not resemble the LIS within $\sim$1~kpc from the sun, even after accounting for solar modulation. 
The differences among the proton spectra obtained by \cite{emissivity2}, by \cite{JM} and by the present work are most likely due to the pion production cross sections and to 
the all-electron spectrum used. 
Indeed, hadronic cross sections are still affected by significant  uncertainties
especially for CRs and target nuclei with atomic number $Z>1$, \citep[e.g.][]{Kamae}
For heavier nuclei the calculated emissivity \citep{JM} have a nuclear enhancement factor of 1.8 for proton-proton interactions as found by \cite{Mori}, while we have 1.5 that would account for  a few per cent difference in the calculation of the emissivity \citep{JM}. 
The best-fit proton spectrum by \cite{emissivity2}, obtained with a sophisticated Bayesian approach with MultiNest, is in agreement with our spectrum down to $\sim$3 GeV. The discrepancy at lower energy is mostly due to differences in the all-electron spectrum used to calculate the bremsstrahlung emissivity component. This bremsstrahlung emissivity component is well constrained by direct CR measurements and synchrotron emission in our present work. \\ 
Similarly to our result on the enhanced proton LIS based on gamma-ray data, an earlier work by \cite{EGB}, which used a different approach still based on propagation models, found the need of increasing the calculated pion-decay emission
component of 50 -- 70 per cent at high energies  
in order to fit {\em Fermi} LAT gamma-ray data at latitudes above 20$^\circ$
up to 500~GeV. However, the main focus of that work was related to obtain the extragalactic 
background emission, hence
the discrepancy between interstellar models and data was
not further investigated. \\
The spectral intensity of the proton spectrum for our baseline best model PDDE is reported in Table 2 in Appendix, together with the proton spectrum that fits the emissivity (Figure~\ref{fig4}).

\begin{figure}
\catcode`\_=12
\includegraphics[width=0.45\textwidth]{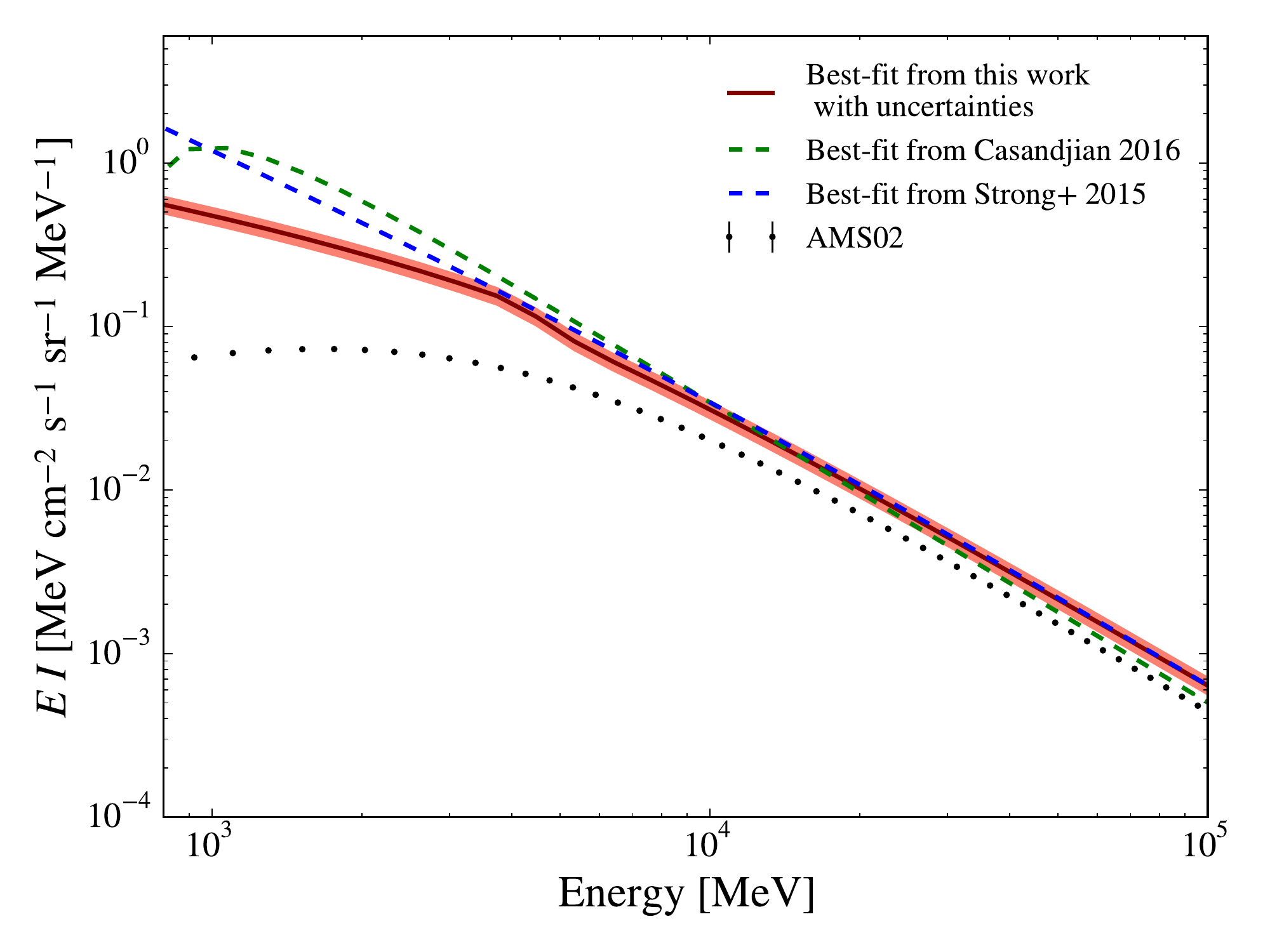}
\caption{Best-fit proton LIS from this work (red solid line) based on emissivity observations \protect\citep{JM} including uncertainties (red region) compared with AMS-02 proton data \protect\citep{AMSpro} (black points). The calculated spectrum is obtained by normalizing the PDDE proton spectrum with the best-fit value of 1.3 (see Table~\ref{Table3}, normalization for the entire energy band). Uncertainties include 1-$\sigma$ error in the normalization and 10 per cent uncertainties in the pion cross section \protect\citep{JM}. Best-fit LIS from \protect\cite{emissivity2} (blue dashed line) and \protect\cite{JM} (green dashed line) are also shown for comparison (uncertainties are not plotted).}
\label{fig4}
\end{figure}

\subsection{Exploring modifications to the baseline models}
In this section we test a modification to one of our models (Section 3.2.1), and we test a different scenario (Section 3.2.2).

\subsubsection{DRELowV model}
\begin{figure}
\catcode`\_=12
\center
\includegraphics[width=0.45\textwidth]{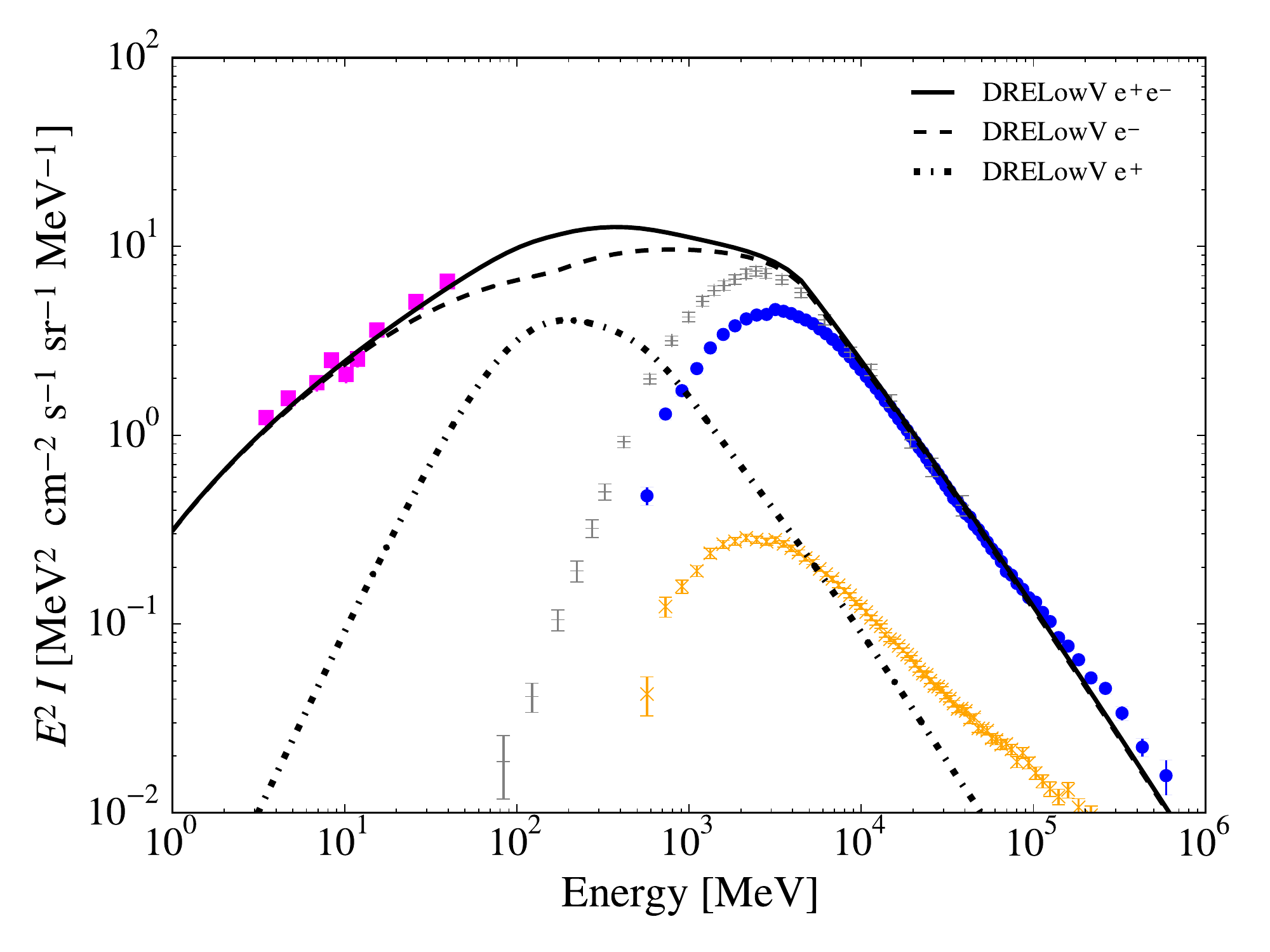}
\caption{Propagated interstellar spectra of the DRELowV model for positrons (dashed-dotted line), electrons only (dashed line), and all-electrons (solid line) compared with data as described in Figure \ref{fig1}.  }
\label{fig8}
\end{figure}

In the effort to find propagation models with reacceleration working both with CR all-electron measurements and with the synchrotron emission we test a modified DRE model.
The modification is based on a recent work \citep{Gulli} where we perform a Bayesian search of the main GALPROP parameters, using the MultiNest nested
sampling algorithm, augmented by the BAMBI neural network machine learning package. 
More specifically, in that work we found that the propagation parameters that best-fit low-mass isotope data (p, p$^-$, and He) are significantly different from those that fit light elements (Be, B, C, N, and O), including the B/C and $^{10}$Be/$^9$Be, secondary-to-primary ratios normally used to calibrate propagation parameters. This suggests that each set of species is probing a different interstellar medium, and that the standard approach of calibrating propagation parameters for all the species using B/C may lead to incorrect results (as previously suggested by the work in \cite{BtoC}). 
Based on this finding, here we explore a different propagation model that we call DRELowV, which represents an attempt to find a reacceleration propagation model that can reproduce CR measurements, and also the synchrotron spectrum as good as the PDDE model. 
In more detail, starting from the DRE baseline model, 
we make some simple modifications to the model parameters in order to reduce the amount of secondary positrons in the range $\sim$(10$^{2}$--10$^{4}$)~MeV, and to consequently better reproduce all the data. 
In particular, we decrease the Alfven velocity of the DRE model to 8.9 km/s for protons and helium only, based on results from \cite{Gulli} previously discussed. We modify the proton spectrum to be similar to the spectra of the baseline models, keeping all the other propagation parameters unchanged. The resulting proton spectrum is shown on Figure~\ref{fig1a}. The spectrum of the light elements are unchanged with respect to the original models, hence, they are not reported here. Spectra and parameters of the light elements can be found in the original paper, where elements up to Si from ACE-CRIS, HEAO3, PAMELA, and CREAM were fitted. 
Then, following the procedure used for the baseline models, here we adjust the electron injection spectral indexes and breaks in such a way that the density of all-electrons in the range $\sim$GeV is similar to the PDDE model. This is now possible because of the lower density of secondaries produced by 
the decreased Alfven velocity with respect to DRE model. Model parameters are summarized in Table~\ref{Table1} and Table~\ref{Table2}.
The resulting DRELowV model requires at least two breaks in the primary injection electron spectrum below a few GeV in order to reproduce the AMS-02 and {\em Voyager 1} data. Figure \ref{fig8} shows the propagated interstellar all-electron spectra for DRELowV model against data. Compared to Figure \ref{fig1} we find that the density of positrons at $\sim$GeV for this model is a factor of 2.5 lower than the baseline DRE and DRC models, and it is similar to the PDDE model.  \\  

\begin{figure}
\catcode`\_=12
\center
\includegraphics[width=0.45\textwidth]{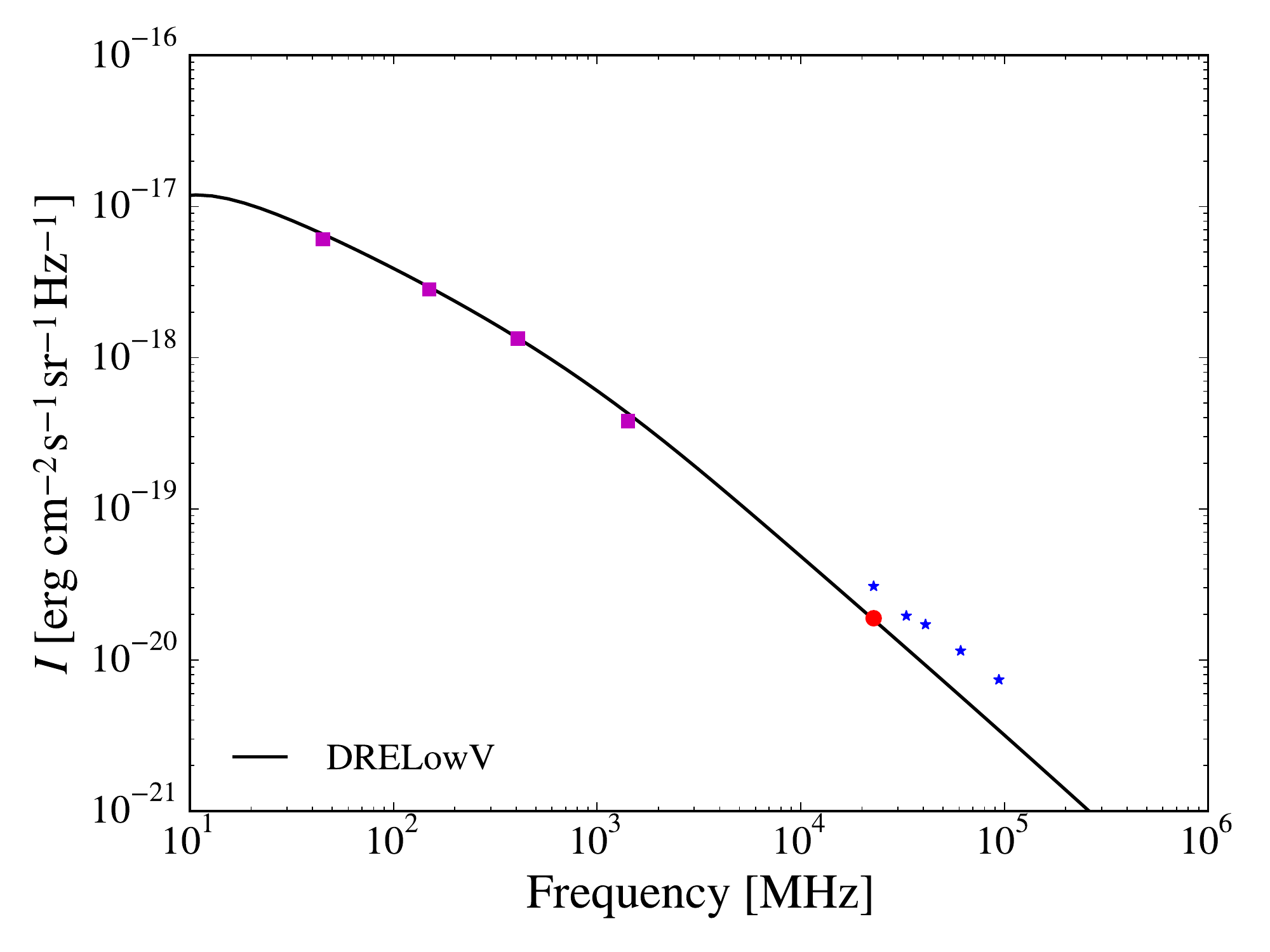}
\caption{Synchrotron spectrum for intermediate latitudes (10$^\circ$$<|b|<$20$^\circ$) of the DRELowV model compared with data as in Figure \ref{fig2}.  }
\label{fig9}
\end{figure}

The synchrotron spectrum is calculated and is shown in Figure~\ref{fig9}. We find that the spectral data are quite well reproduced in the whole frequency range, as for the case of PDDE model.  
As a consequence this propagation model and the resulting LIS are a good
representation  of the spectrum  that produces  the synchrotron  emission, as found for PDDE model. This suggests that the contribution of secondaries and primary electrons is now well constrained, meaning that it is possible to find a propagation model with reacceleration (with significantly reduced reacceleration compared to the usually assumed for protons) consistent also with radio synchrotron data.\\

\begin{figure}
\catcode`\_=12
\center
\includegraphics[width=0.4\textwidth]{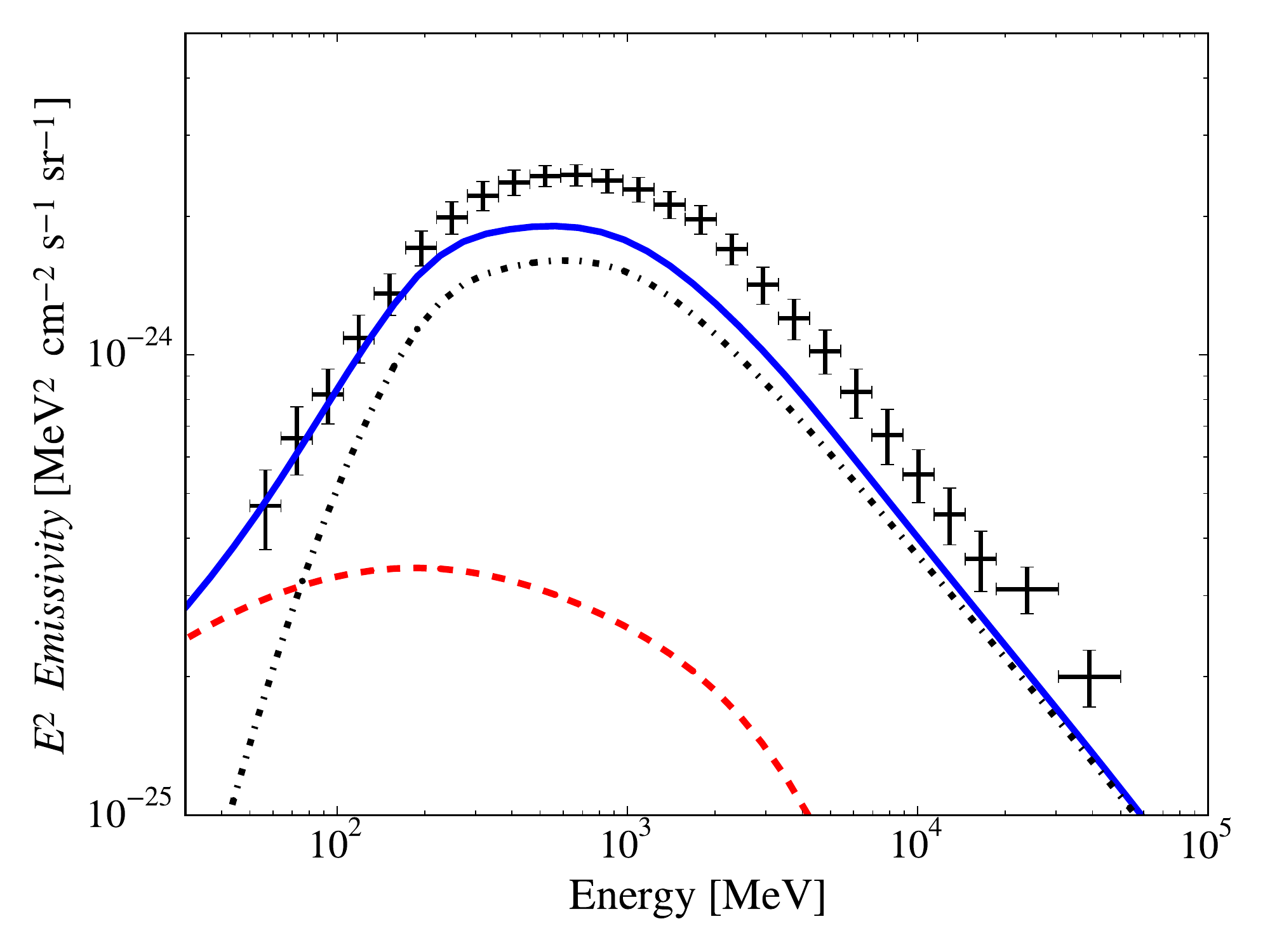}
\includegraphics[width=0.4\textwidth]{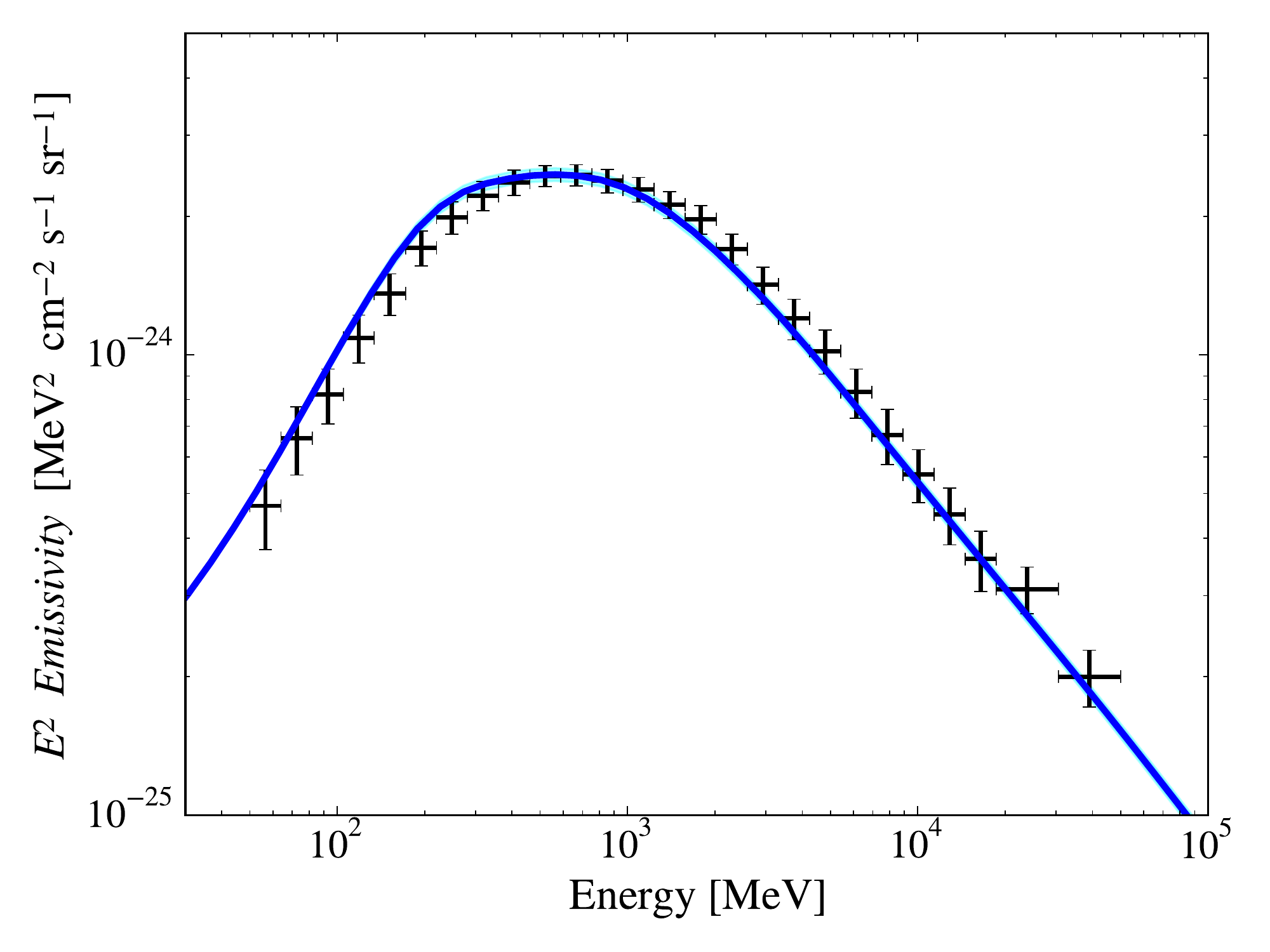}
\caption{Local gamma-ray emissivity of DRELowV model compared with {\em Fermi} LAT local HI emissivity \protect\citep{JM}. {\em Top}: Calculated bremsstrahlung component (red dashed lines), pion-decay component (blue dashed-dotted lines), and their sum (blue lines) are shown.  {\em Bottom}: the sum (blue lines) of the calculated bremsstrahlung and the fitted pion-decay component, and the 1-$\sigma$ error (cyan region) in the fitted parameter are shown. The fitted parameter for the pion decay component, errors, and chi-squares of the fit are reported in Table~\ref{Table3}.} 
\label{fig10}
\end{figure}

Following the same procedure as used for the baseline models, we calculate the local gamma-ray emissivity for DRELowV model and we compare it with data. Figure \ref{fig10} shows that, similarly to what happens to the PDDE model, a very good agreement is visible for the DRELowV model below a few hundred MeV (top plot). 
This confirms the preference of models with low all-electron density at the $\sim$GeV range. 
At higher energies, instead, the predictions of the local gamma-ray emissivity 
are still $\sim$~30~--~40 per cent lower than the {\em Fermi} LAT observations, as also found for the baseline models. This suggests that proton CR measurements are not resembling the LIS within $\sim$1~kpc even if accounting for solar modulation, as found in Section 3.1. 
Figure~\ref{fig10} (bottom plot) shows the normalized emissivity, while Table~\ref{Table3} summarizes the best-fit results for this model, leading to scaling factors very similar to the PDDE model. This is not surprising since the all-electron LIS and the proton LIS of the two models are alike. \\ 
Note that in principle modifications to the DRC model as performed for the DRELowV model could be possible. However repeating the procedure as in \cite{Gulli} to obtain a fully Bayesian parameter estimation for the DRC model including convection is beyond the present effort. 

\subsubsection{The electron LIS at high energies}
In the following we aim at verifying whether our initial assumption on the 'positron excess' affects the results.
We assume here that the high-energy positron spectrum (that includes the 'positron excess') is produced by injection and propagation and it is not peculiar to our position in the Galaxy and our proximity to an electron-positron source. 
In other words we assume the distribution of the Galactic sources producing positrons at high energies to be the same compared to the distribution of the sources of primary electrons. 
Once computed the synchrotron emission we find that this modification does not effect the intensity in radio and microwaves, i.e. radio and microwaves are not sensitive to the energy range of the 'positron excess'. In addition, we also find that this modification does not effect the computed gamma-ray emissivity either, because electrons and positrons are too energetic to contribute to the emission. 
Consequently, neither radio/microwaves nor the gamma-ray emissivity is affected by positrons at those energies, which could instead contribute to the gamma-ray emission at high energies via IC above a few GeV. 

\begin{figure*}
\center
\catcode`\_=12
\includegraphics[width=0.4\textwidth]{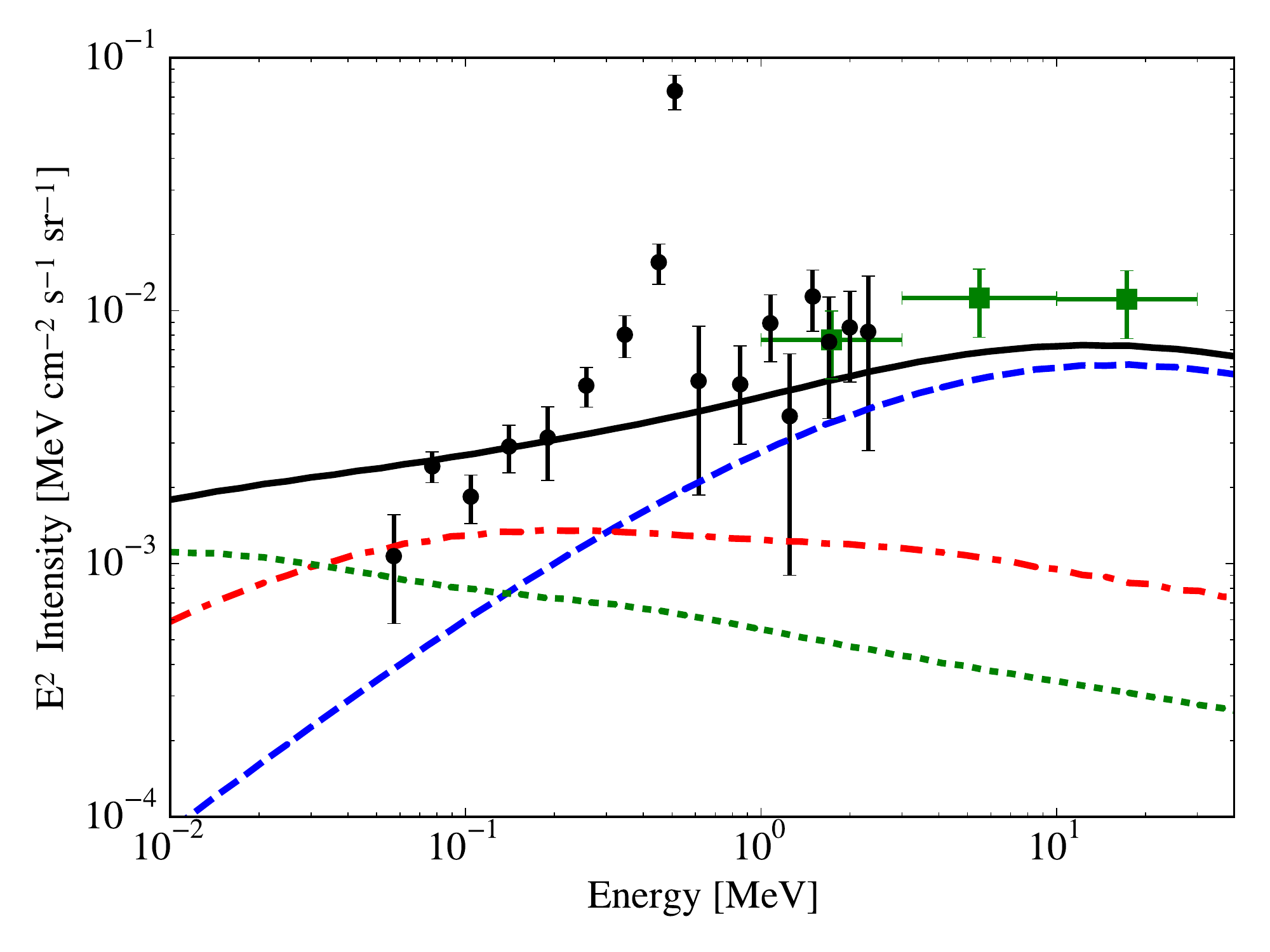}
\includegraphics[width=0.4\textwidth]{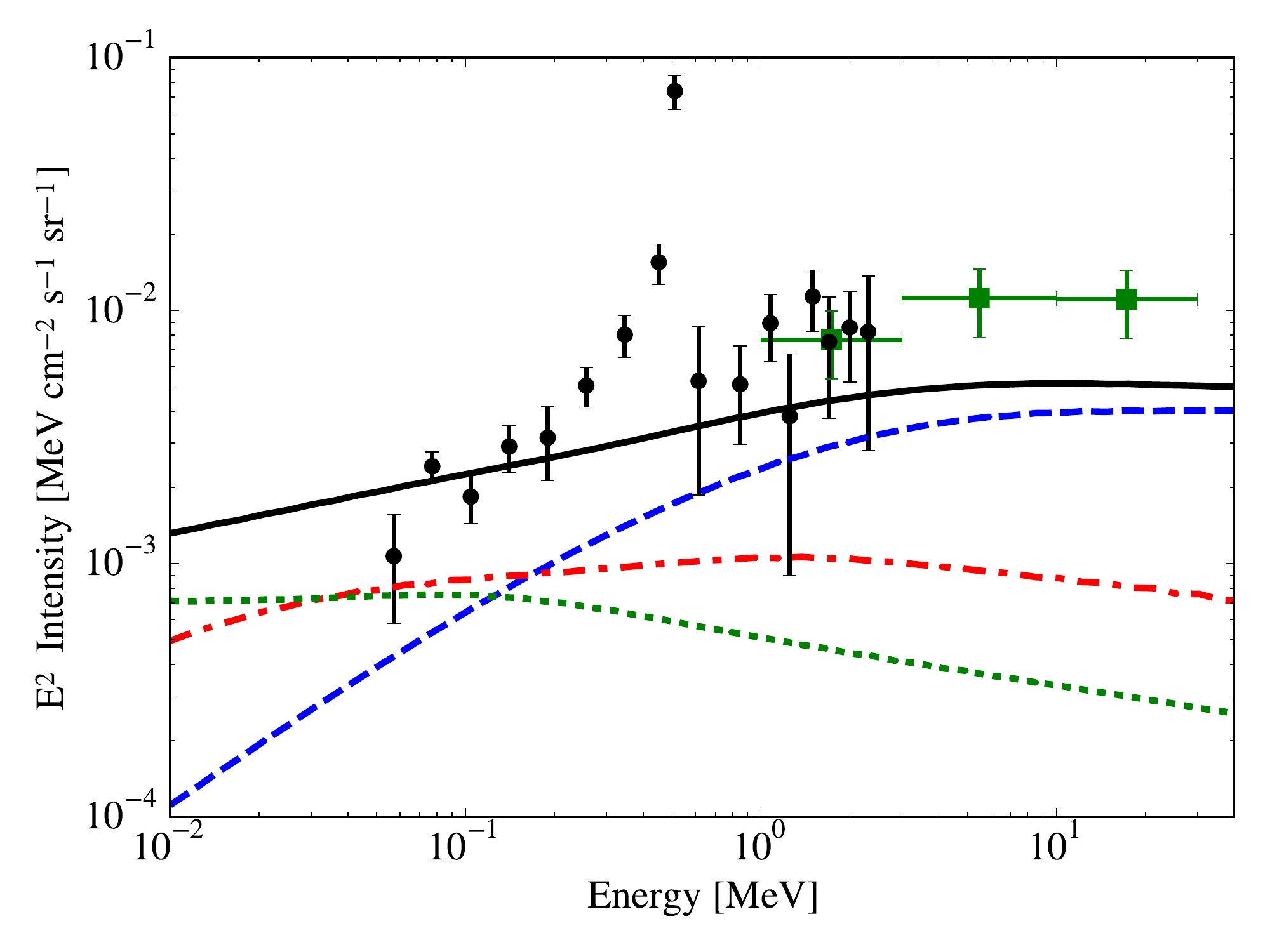}\\
\includegraphics[width=0.4\textwidth]{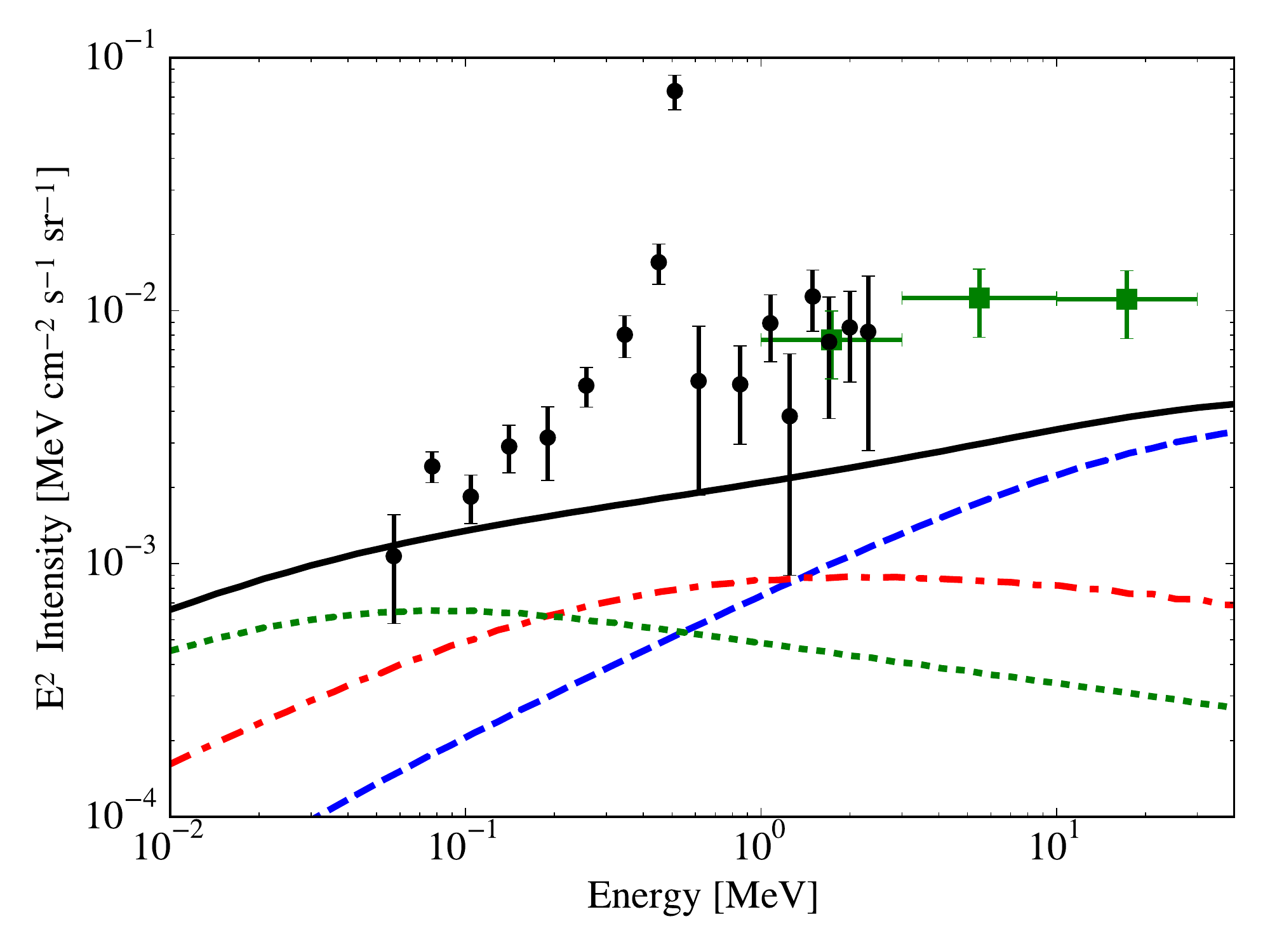}
\includegraphics[width=0.4\textwidth]{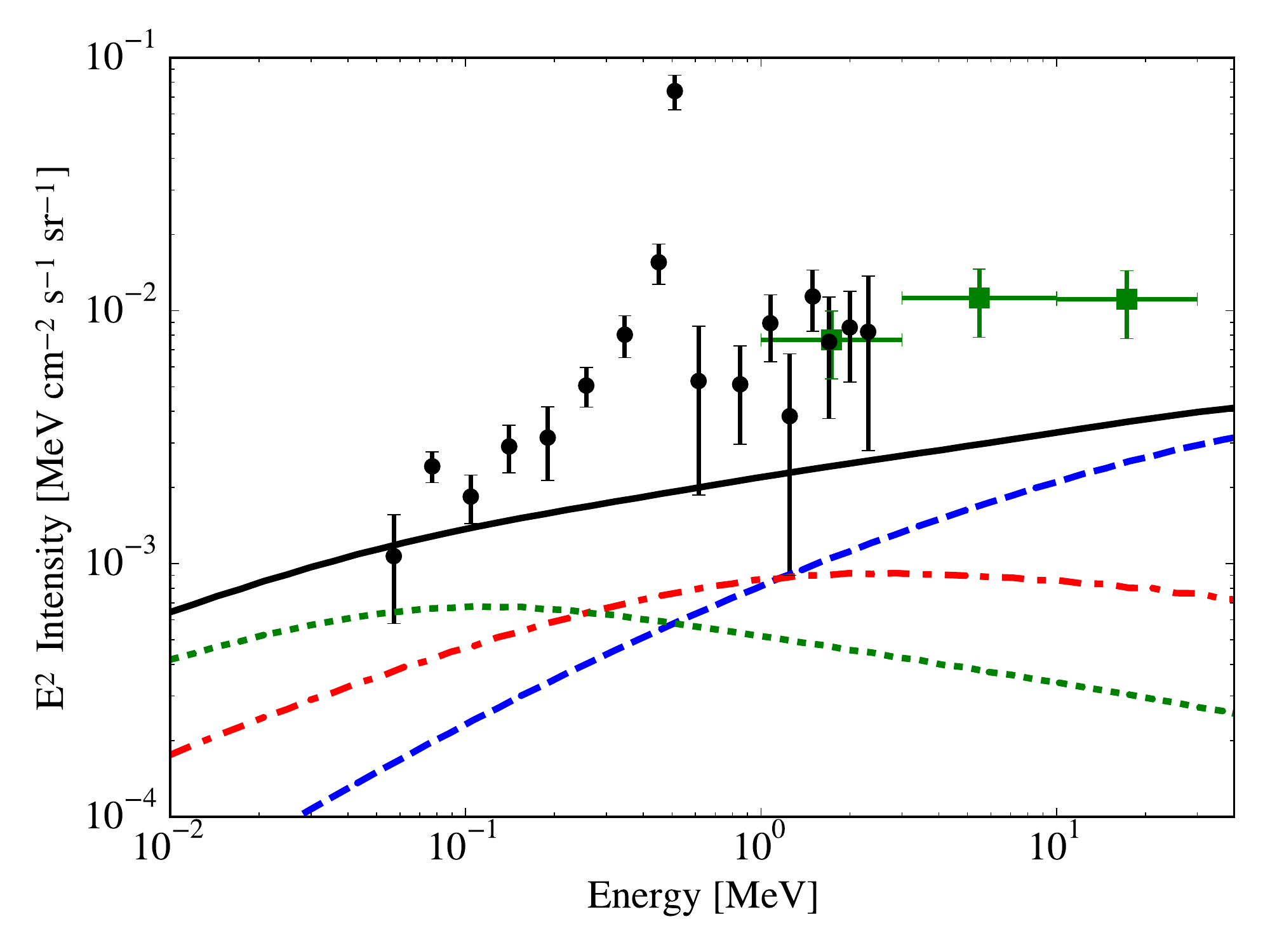}\\
\includegraphics[width=0.4\textwidth]{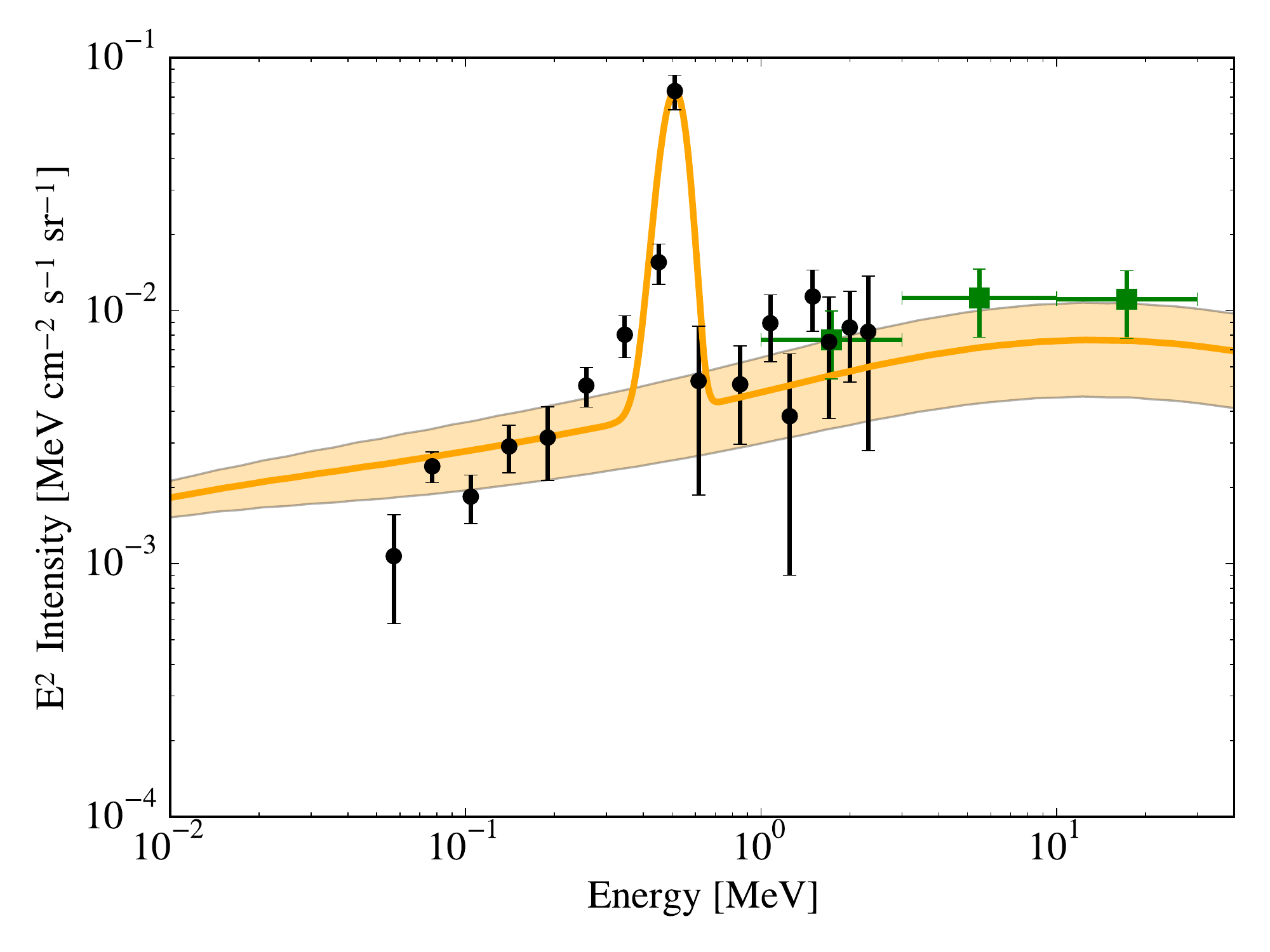}
\includegraphics[width=0.4\textwidth]{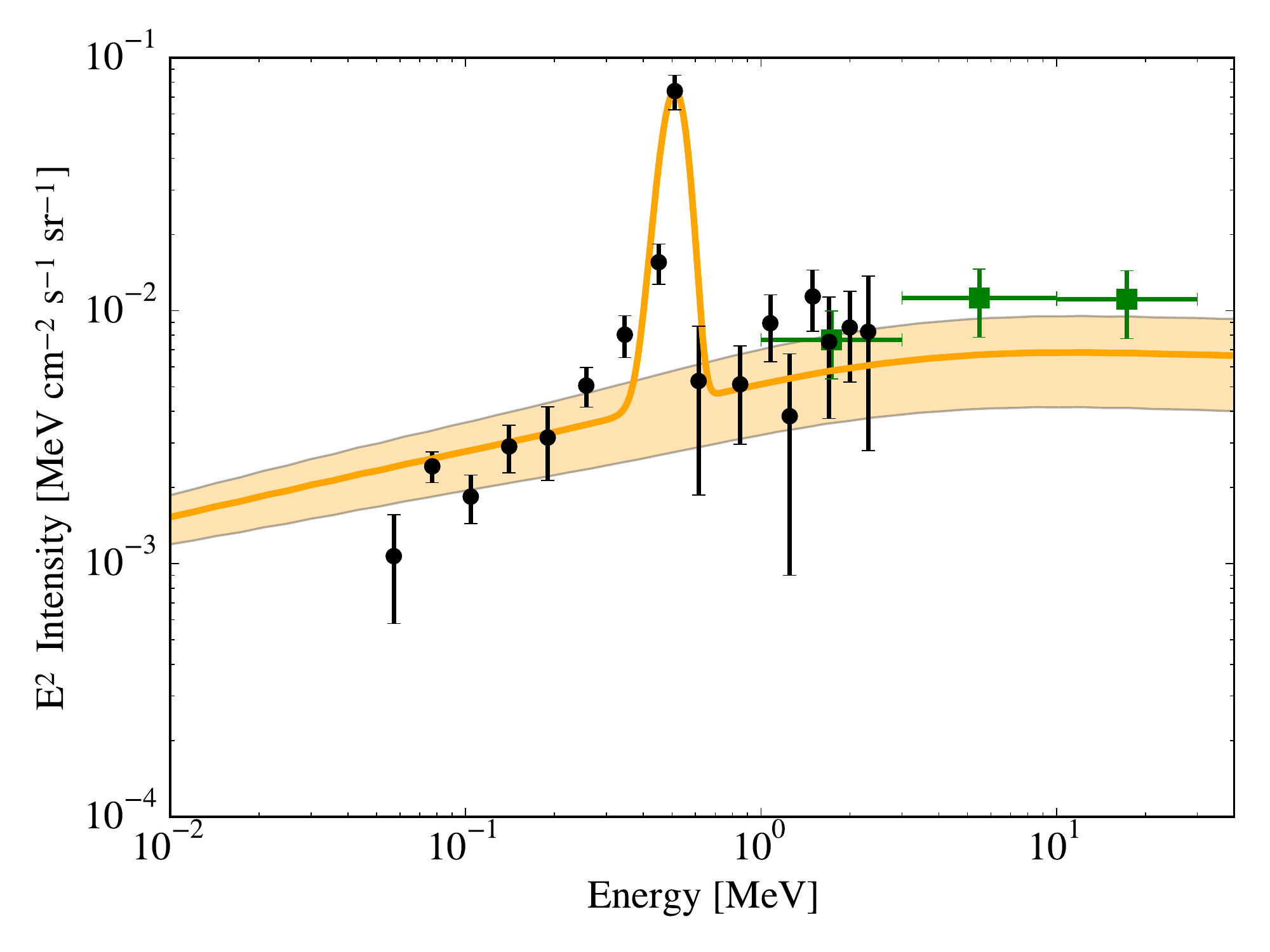}\\
\includegraphics[width=0.4\textwidth]{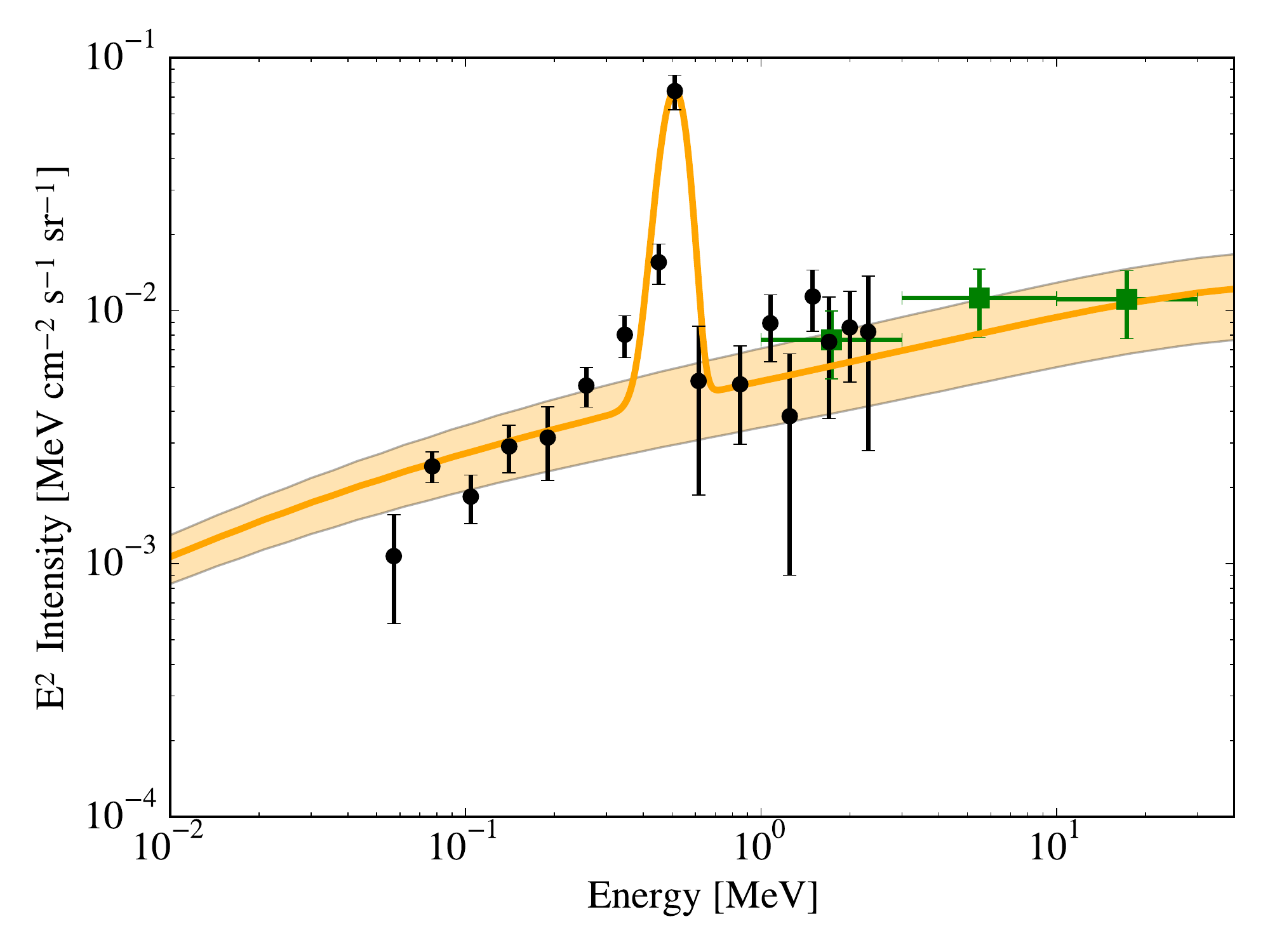}
\includegraphics[width=0.4\textwidth]{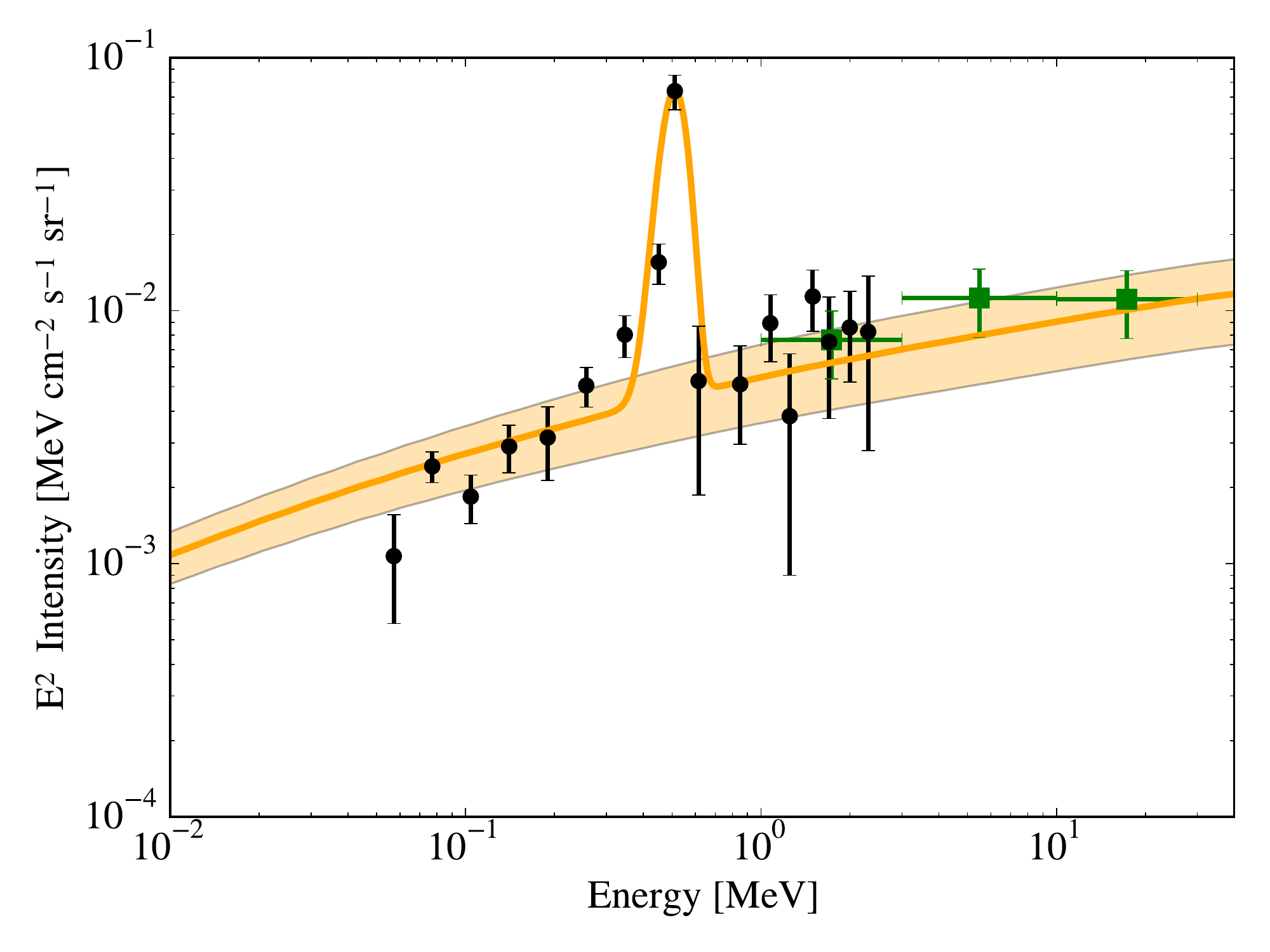}
\caption{X-ray and soft gamma-ray spectra of the models, left to right, DRE, DRC, PDDE, and DRELowV. Top two rows show the three IC components as modeled: on the CMB (green dotted line), on the diffuse IR (red, dash-dotted line), and on the diffuse optical (blue dashed line), along with their sum (black solid line). Bottom two rows show the sum of the components after the fit (orange line) with one sigma error region (orange region). Data are from \protect\cite{Bouchet} for the inner Galaxy $|b|<$15$^{\circ}$ and $|l|<$30$^{\circ}$ from SPI (black points) and from COMPTEL (green points) with respective error bars. Fit results are reported in Table~\ref{Table4}.}
\label{fig17}
\end{figure*}

\subsection{X-rays and soft gamma rays from the inner Galaxy}
After studying the spectra of CRs in the local interstellar medium,  we use our resulting models, DRE, DRC, PDDE, and DRELowV, to compute the emission from the inner Galaxy observed in the range 0.1 - 30 MeV, following the work in \cite{Bouchet} and in \cite{Porter2008}, to see how they compare to X-rays and soft gamma-ray data. 
Our sky region of interest is $|b|<$15$^\circ$ and 330$^\circ$$<l<$ 30$^\circ$.
In this energy range IC emission is the only CR-induced interstellar component. We separately calculate the contributions to the IC intensity by optical, infrared (IR), and CMB photons.
Figure~\ref{fig17} shows the spectral component contributions to the diffuse IC emission for all the models (two upper rows of the figure, left to right: DRE, DRC, PDDE, and DRELowV) compared to SPI and COMPTEL spectral data, as published by \cite{Bouchet}. 
The three IC components (by CMB, by optical, and by IR photons) are visualized, together with their summed emission. 
To account for uncertainties in the ISRF we fit the normalization of the IC components to the data with the following method. 
Because the optical and IR components are physically related, a
common scaling parameter for both is used following the work in \cite{diffuse2} by the {\em Fermi} LAT Collaboration. The CMB component is instead fixed since the CMB is known. 
A gaussian emission line at 511 keV for the electron-positron annihilation is also added. The best-fit values for all the models are collected in Table~\ref{Table4}, while the resulting fitted IC emission is shown in Figure \ref{fig17} (two bottom rows, left to right: DRE, DRC, PDDE, and DRELowV model).
We find that while our preferred PDDE and PDDELowV models require a scaling factor of $\sim$3 in the optical and IR components in order to reproduce the data, for the DRE model the spectral shape and intensity of the diffuse IC emission matches reasonably well the data. 
Overall, the DRE and the DRC models reproduce the intensity of the data by
SPI and by COMPTEL better than the PDDE and the DRELowV models.
Their higher IC intensity with respect to PDDE and DRELowV models is due to the enhanced all-electron spectrum of those models in the $\sim$(10$^{2}$--10$^{4}$)~MeV range. 
This is reflected in the best-fit scaling factors found to be $\sim$1 and $\sim$1.3 for DRE and DRC models respectively, as reported in Table~\ref{Table4}. 
In general we find that models with the all-electron  LIS that fit the local synchrotron emission and the local emissivity (PDDE and DRELowV) underestimate the X-ray emission in the inner Galaxy. Instead, a significant contribution from secondary positrons and electrons (as in DRE and DRC models) reproduces observations by SPI and COMPTEL of the inner Galaxy without the need of substantially enhancing the ISRF.

\begin{table}
\begin{center}
\caption{Best-fit values of the IR and optical component of the IC emission derived from comparison of X-ray and soft-gamma observations by INTEGRAL/SPI and COMPTEL.}
\begin{tabular}{lccc}
\\
 \hline
 \hline
 \\
 {Model} &  {IR/Optical normalization} &   {chi-square}               
  \\
 \hline
\\
DRE & 1.05 $\pm$~0.44 &   2.92  \\
DRC & 1.35 $\pm$~0.55 &  2.76 \\
PDDE & 2.98 $\pm$~1.13 &  2.42 \\
DRELowV & 2.95 $\pm$~1.11 & 2.37\\
\\
\hline
\\
\label{Table4}
\end{tabular}
\end{center}
\end{table}

\subsection{Gamma rays from the Galactic centre}

\begin{figure*}
\center
\catcode`\_=12
\includegraphics[width=0.45\textwidth]{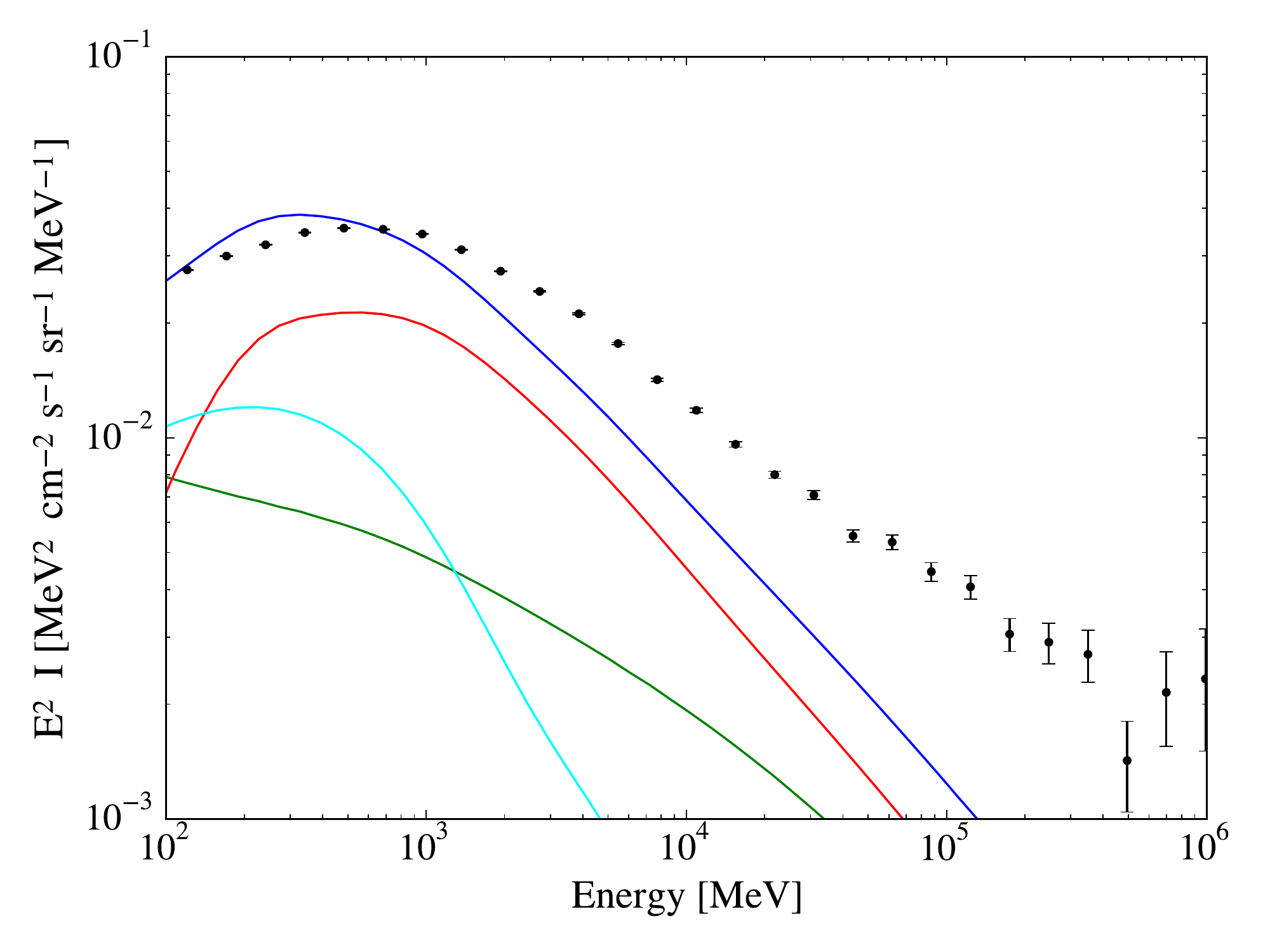} 
\includegraphics[width=0.45\textwidth]{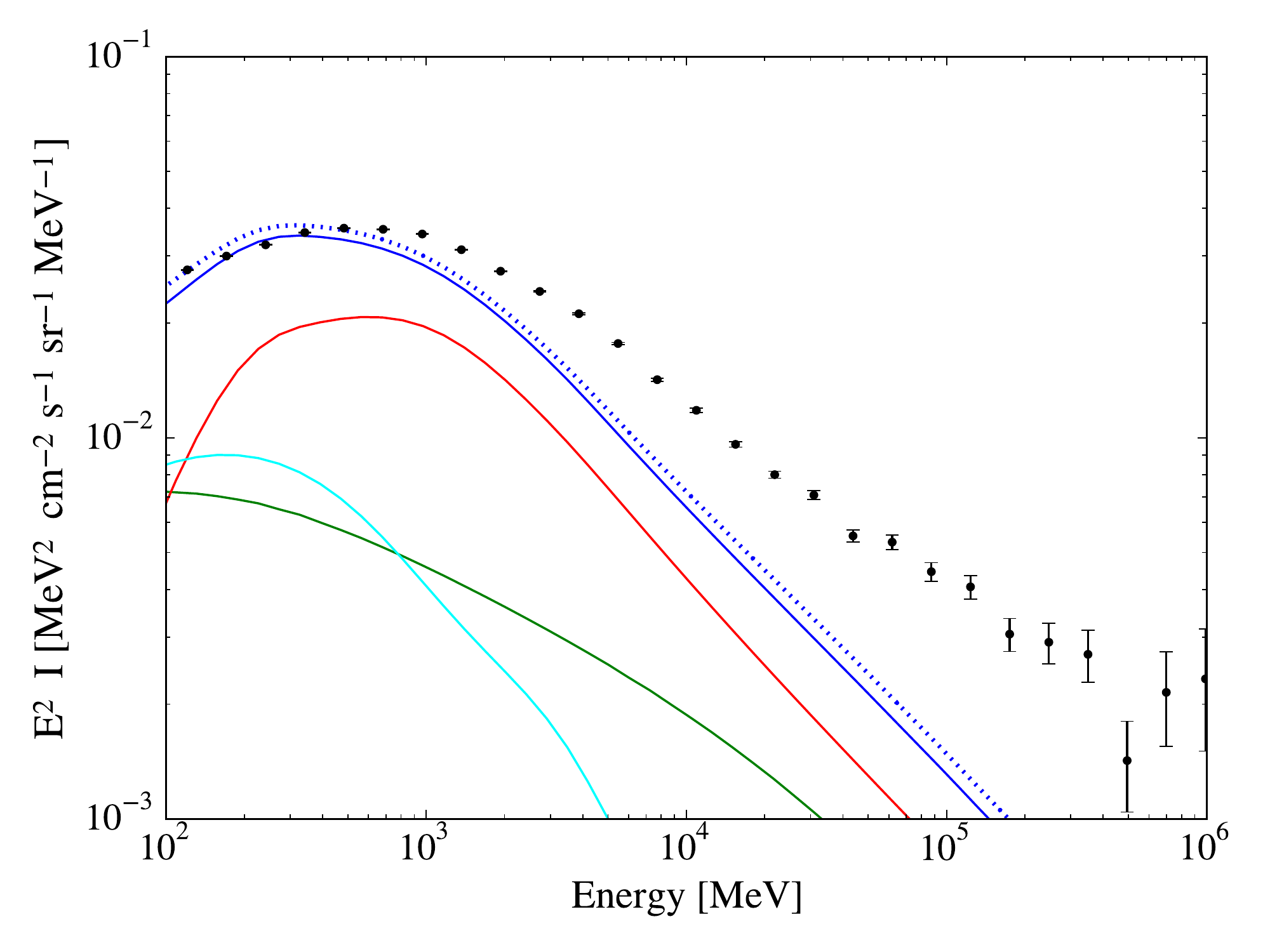} \\
\includegraphics[width=0.45\textwidth]{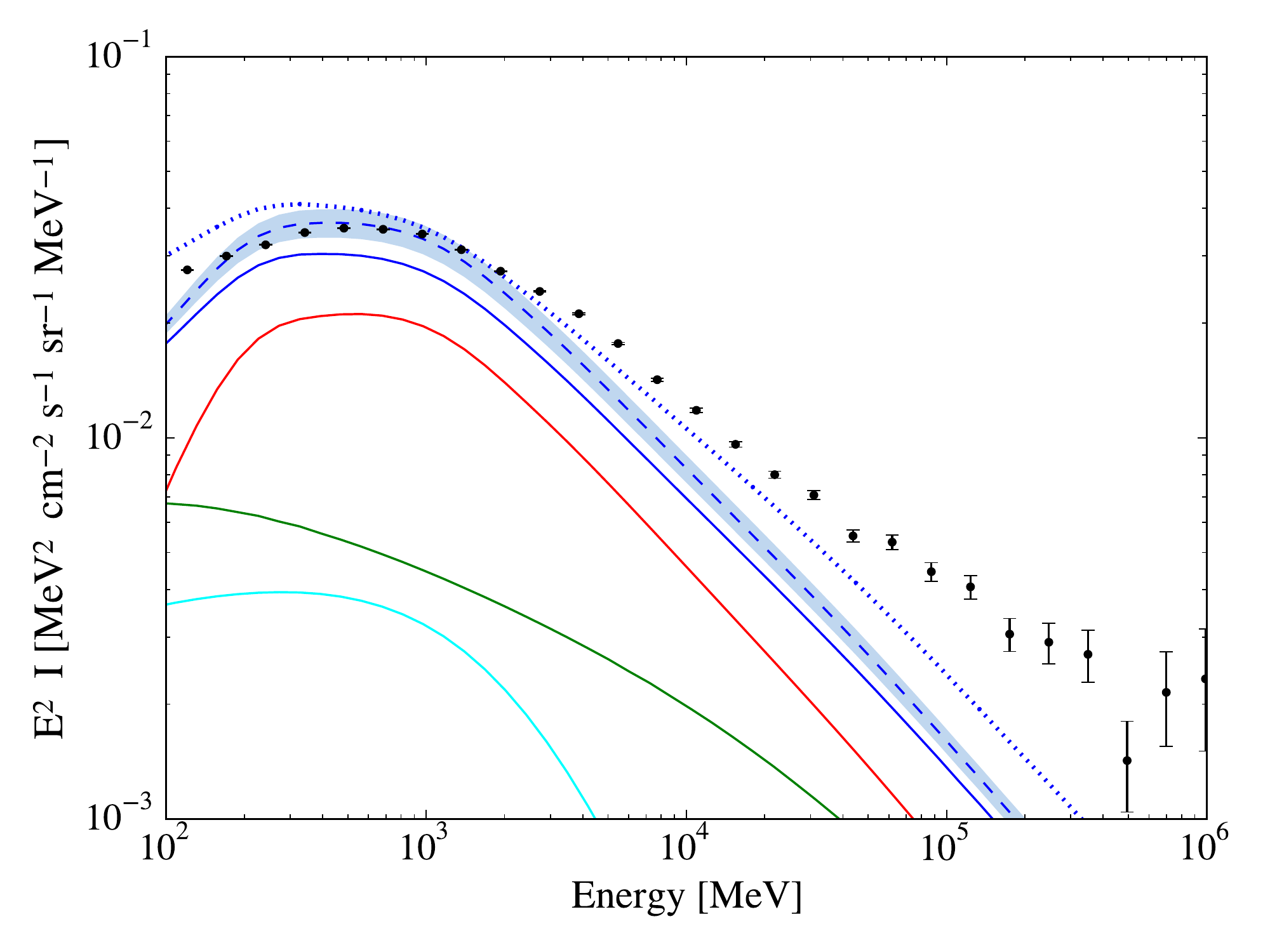}
\includegraphics[width=0.45\textwidth]{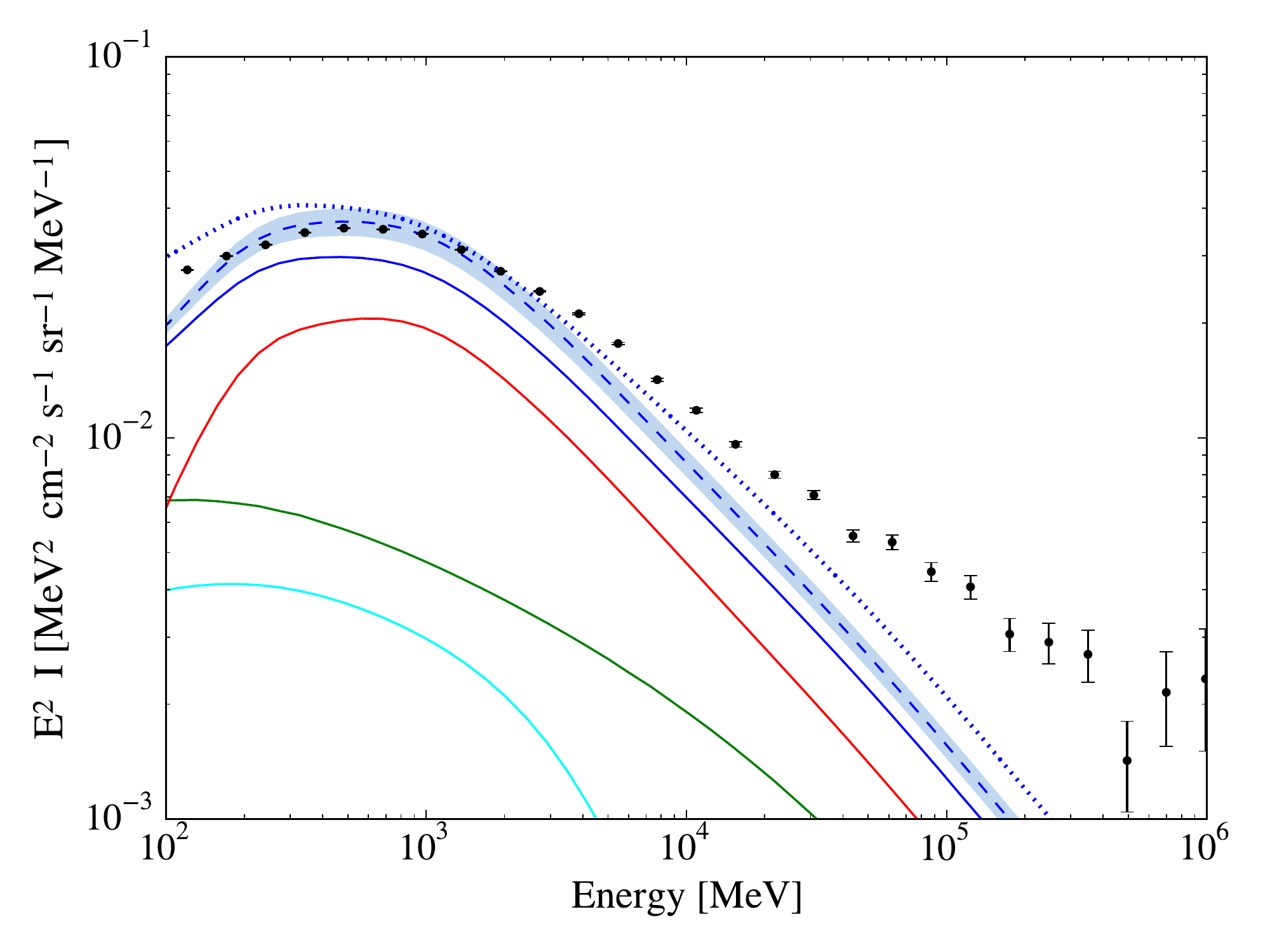}
\caption{Gamma-ray spectral intensity for the inner Galaxy (10$^\circ$ radius around the Galaxy centre) of the four propagation models, left to right top to bottom, DRE, DRC, PDDE, and DRELowV. Data (black points) are {\em Fermi} LAT spectra for the region 10$^\circ$ radius around the Galactic centre from \protect\cite{P8IG}. 
The calculated total interstellar emission (blue solid lines) is the sum of the bremsstrahlung (cyan line), the IC (green line), and the pion-decay (red line) components for the propagation models that reproduce CR measurements. The same models with proton spectra based on the gamma-ray emissivity (blue dashed lines with blue-grey shaded region) with normalizations from Table~\ref{Table3} for the entire energy band, and with normalized ISRF based on SPI and COMPTEL data (blue dotted lines) with normalizations from Table~\ref{Table4} are also shown. 
The most luminous sources in this regions are masked. Models are treated in the same way as data. Additional emission components are not plotted, such as isotropic, faint sources, GeV excess, solar and lunar emission. Models are not fitted to data. DRE and DRC models do not fit the emissivity below $\sim$0.4 GeV, hence the models with normalized proton spectrum are not shown.} 
\label{fig16}
\end{figure*}

Over the last years the Galactic centre has become a region of particular interest
to the astrophysical community. Especially at gamma-ray energies, the properties of
this sky region might encode possible discoveries \citep[e.g.][]{IG1, Calore, Carlson, IG2, P7IG, P8IG}. Therefore, any effort in modeling the emission in this region is important. \\
Studies in this region often fit the interstellar model components (IC, pion, and bremsstrahlung) to data in bin-by-bin of energies. 
Being fitted bin-by-bin the information of the CR spectra is lost because each bin is independently adjusted together with other components (i.e. detected sources, isotropic emission, solar and lunar emission). Our approach is instead to directly compare our propagation model (DRE, DRC, PDDE, DRELowV) with {\em Fermi} LAT data with no spectral adjustments, and no fit to data. 
This is useful for illustration and for investigating whether present observations in this region allow
for challenging some models without performing a dedicated analysis that would account for all the emission components in this difficult region. 
In fact, if the sum of the components (pion, bremsstrahlung, and IC) of one of our propagation models overestimates the data, it means that this model needs more attention.
Moreover, while we discuss the comparison of interstellar models with data, we do not draw any new final conclusion by looking at this region alone, which would need a dedicated work.
We compare our propagation models with the {\em Fermi} LAT spectral data over an area of 10$^\circ$ radius around the Galactic centre taken from a very recent study by \cite{P8IG}. 
The comparison of models with data is shown in Figure~\ref{fig16}, where
each plot represents one model at a time (top to bottom, left to right DRE, DRC, PDDE, and DRELowV). 
We plot the gamma-ray intensities due to the bremsstrahlung (cyan solid lines), the IC (green solid lines), the pion-decay (red solid lines), and their sum (blue solid lines) for the propagation models that reproduce CR measurements. 
For DRE model at energies below 1 GeV our computed total (sum of bremsstrahlung, pion decay, and IC) interstellar emission alone (blue solid lines) over-predicts the {\em Fermi} LAT data (black points). The summed component for the DRC model is accepted by the data, if no other components (e.g. additional sources below $\sim$1~GeV ) are included\footnote{The other components of the gamma-ray emission seen by {\it Fermi} LAT are not shown (i.e. isotropic, faint sources, solar and lunar emission, etc.), because this would need a dedicated work, which is beyond the present effort.}. 
While baseline DRE and DRC models may be challenged by gamma-ray data in this region, the PDDE
and DRELowV models (two plots in second row of
Figure~\ref{fig16}, blue solid lines) provide a better spectral representation of the {\em Fermi} LAT data below 1~GeV (blue solid lines). Being the pion decay emission produced by similar hadronic CR spectra for all the models,
the major contribution to this difference among the models is given by the bremsstrahlung component, due to the different electrons and positrons.
In addition, for all the models, the resulting components with normalized ISRF as found to fit the SPI and COMPTEL data in the inner Galaxy and reported in Table~\ref{Table4} are also plotted (blue dotted lines). Moreover, models with proton spectra scaled with the best-fit normalization in Table~\ref{Table3} (for the entire energy band), which are based on the local gamma-ray emissivity data, are shown for PDDE and DRELowV models\footnote{DRE and DRC models do not fit the emissivity below $\sim$0.4 GeV} (blue dashed lines, with blue-grey shaded region).
The plots shows that PDDE and DRELowV models with enhanced ISRF and proton spectrum may be challenged as well below $\sim$~1~GeV once other components (i.e. isotropic, faint sources, solar and lunar emission) are included. 
In fact, for the PDDE and DRELowV models, an increase of the ISRF (blue dotted line) 
would imply also an enhancement of the IC emission below a few GeV. The need for an increase of the IC emission component in the Galactic centre region
was claimed in a recent study by \cite{P7IG}, but the degeneracy between ISRF and electrons was not solved. 
However, in that analysis energies below 1 GeV were not included. By extending to energies down to
100 MeV, our comparison may suggest that an enhanced ISRF could be disfavored, favoring the alternative hypothesis of a harder electron spectrum in that region only.

\subsection{Implications on the results from possible additional uncertainties in the data}

In this work we show the feasibility and importance of using multiwavelength observations, together with CR measurements, to study the LIS and propagation models. Here we discuss possible uncertainties in the data and the implications to our results. 

Regarding the study of the synchrotron emission, the exact derivation of the synchrotron maps as obtained by {\em Planck} and {\em WMAP} have limitations, due to the various assumptions required and degeneracies in separating multiple astrophysical components including synchrotron, free-free, thermal dust and anomalous microwave emissions (AME) \citep{LFPlanck}. 
As a consequence there are likely degeneracies among the various low-frequency components, especially between AME and synchrotron in the Galactic plane. 
While the {\em WMAP} synchrotron intensity is clearly overestimated, the {\em Planck} synchrotron intensity may be slightly underestimated \citep{LFPlanck}. As a direct consequence it is clear from Figure~\ref{fig2} that possible uncertainties would not change our conclusion on the preference of PDDE model over the DRE and DRC models based on radio and microwave data. \\
It is worth noting that also the zero levels of the radio surveys are not clearly determined. The detailed work of \cite{Wehus2017} estimated a monopole of 8.9$\pm$1.3~K in the 408 MHz map, which includes any isotropic component (CMB, Galactic and extragalactic), which we use for the fit.
In our previous work \citep{Strong2011} we adopted a 3.6~K offset, which slightly increased the excess at lower frequencies for the diffusive-reacceleration models. 
Moreover, as discussed above, intermediate latitudes are not significantly affected by the choice of the offset. In addition a much larger offset in the radio surveys would lead to an even larger discrepancy between data and the DRE and DRC models. 
This would also affect the PDDE and the DRELowV models, yet to a much smaller extent compared to the DRE and the DRC models.
Further model-dependent studies and data from MHz to tens of GHz, including the Square Kilometre Array telescope \citep[e.g.][]{Clive} and C-BASS \citep{CBASS}  
will help in separating the components and may provide more stringent constraints to the all-electron  spectrum. Future observations could also help in explaining the isotropic radio excess seen for example by ARCADE 2 \citep{Arcade}. \\
The gamma-ray HI emissivity is an important indirect observable of CRs. 
Uncertainties in its extraction from the {\em Fermi} LAT data may come from 
the lack of precise knowledge about the gas column densities, including
gas not traced by HI or CO. Indeed, even though the emissivity derivation is given for atomic
hydrogen that is well traced by the 21-cm line, possible correlations between the gas phases might not allow for a full
separation of the components. Another uncertainty related to the gas comes from the HI spin temperature assumed to correct for the opacity. This issue has been most likely addressed in \cite{JM}, in which
different spin temperatures are tested assuming a constant spin temperature in the Galaxy.  \\

\section{Importance of the future missions e-ASTROGAM and AMEGO}
\begin{figure}
\catcode`\_=12
\includegraphics[width=0.471\textwidth]{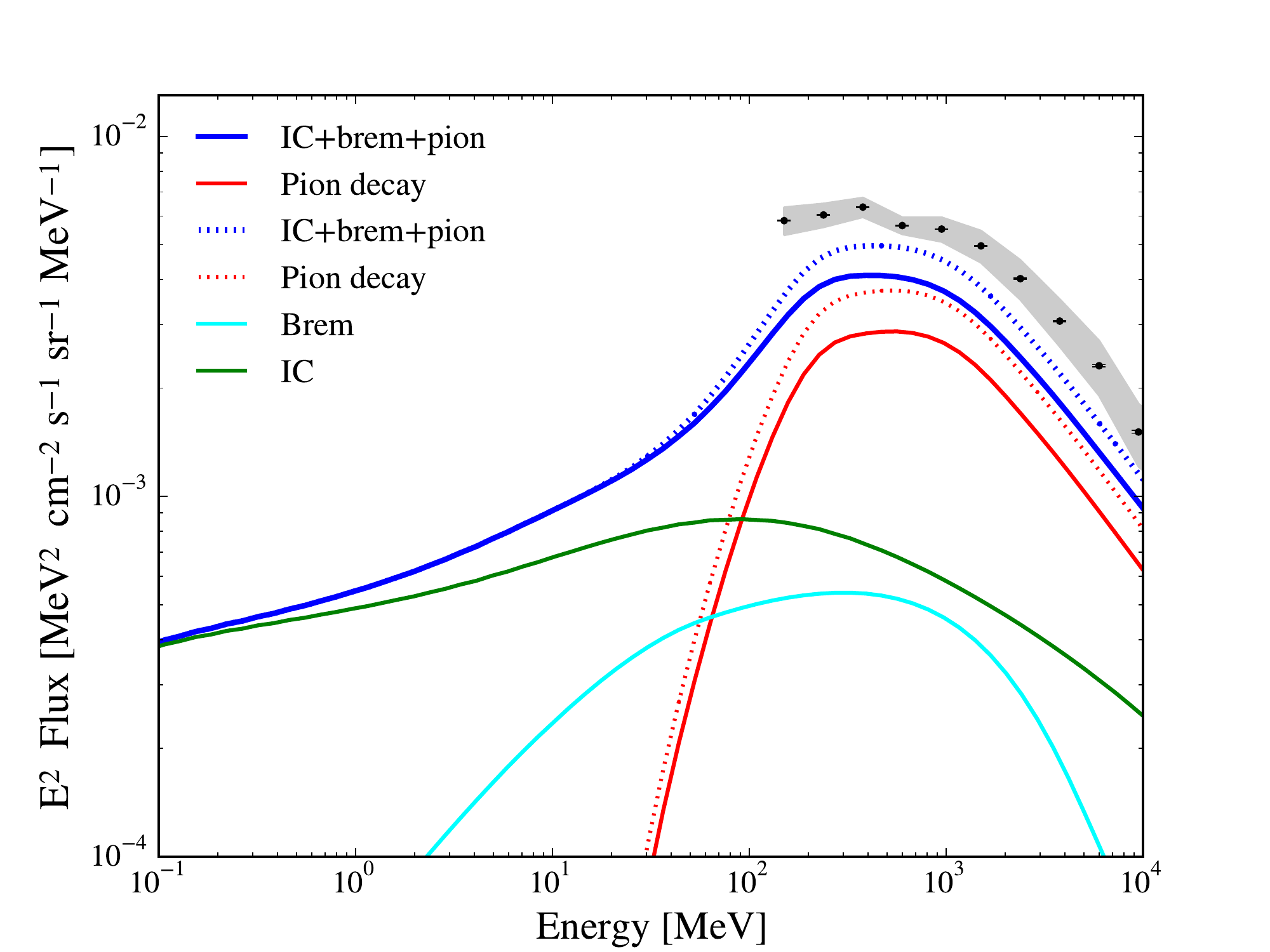}
\includegraphics[width=0.45\textwidth]{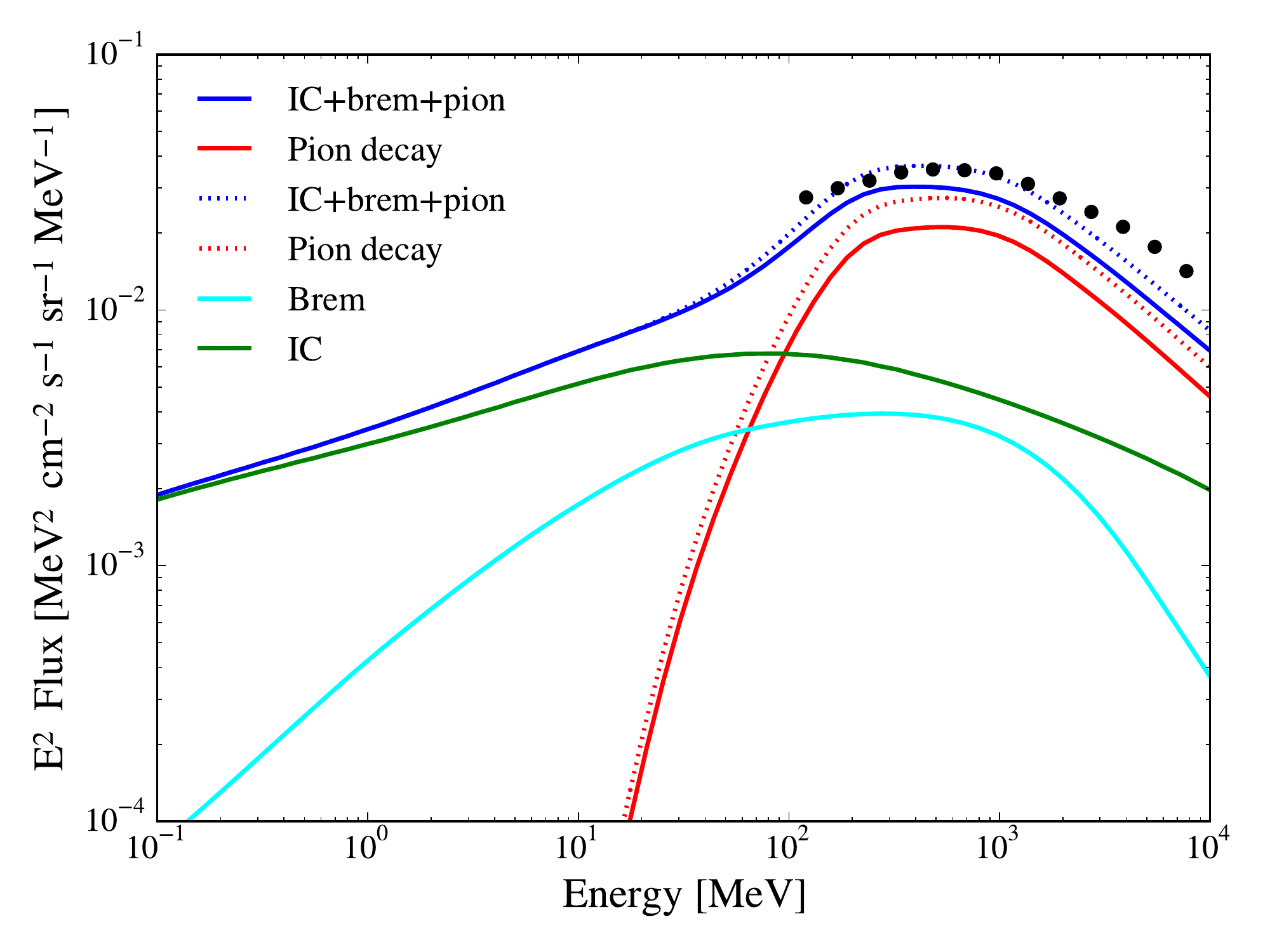}
\caption{Predictions of the interstellar emission for the energy range of e-ASTROGAM and AMEGO gamma-ray instruments for PDDE model. {\em Top}: intermediate latitudes (10$^\circ$$<|b|<$20$^\circ$) for baseline PDDE model (solid lines) and for PDDE model with enhanced protons (dashed lines) that reproduces the gamma-ray emissivity.  The different components are pion decay (red), bremsstrahlung (cyan), IC (green), and total interstellar (blue).
Data are {\em Fermi} LAT spectra for intermediate regions from \protect\cite{diffuse2} (black points). Data include statistical (grey area) and systematic errors (black bars).  
{\em Bottom}: predictions for the inner Galaxy (10$^\circ$ radius around the Galaxy centre) for baseline PDDE model (solid lines) and for PDDE model with enhanced protons (dashed lines). The different components are as in the top figure. 
The e-ASTROGAM extended-source sensitivity is below the plotted intensity (see text for more details). Other components of the gamma-ray sky are not plotted.}
\label{fig18}
\end{figure}

The {\em Compton Gamma-Ray Observatory} with its COMPTEL telescope \citep{Comptel} has explored the MeV band
to the best sensitivity as of today.  
The COMPTEL Catalog
\citep{ComptelCat} contains 32 steady objects. 
The newly proposed MeV missions require accurate astrophysical diffuse background models to detect sources on the MeV sky. 
More precisely, e-ASTROGAM \citep[enhanced ASTROGAM,][]{eastrogam} is designed to detect gamma rays from 0.3~MeV
to 3~GeV. The proposed AMEGO mission\footnote{https://asd.gsfc.nasa.gov/amego/index.html} (the All-sky Medium Energy
Gamma-ray Observatory) covers a very similar energy band from 0.2~MeV to 10~GeV. 
In Figure~\ref{fig18} we extend our best model (PDDE)  
down to 0.1~MeV, and we predict the diffuse interstellar emission 
at intermediate latitudes (10$^\circ$$<|b|<$20$^\circ$, upper panel) and in the Galactic centre region ($10^\circ$ radius, lower panel).  
Plots show our baseline PDDE model (solid lines), and the PDDE model with enhanced proton spectrum that fits the gamma-ray emissivity (dashed lines, scaled with the best-fit normalization in Table~\ref{Table3} for the entire energy band). A major uncertainty comes from the adopted proton LIS, affecting predictions at energies above $\sim$100~MeV  where the pion decay component is dominant. Predictions at $\sim$MeV energy range for PDDE model are not significantly affected by the enhanced hadronic spectrum, due to the dominance of leptonic components. In fact, the all-electron spectrum has been well constrained in this work by both CR direct measurements and synchrotron data. 
The e-ASTROGAM extended-source sensitivity for one year of observations based on simulations for the inner Galaxy is below the plotted intensity, being of the order of a few 10$^{-5}$ cm$^{-2}$s$^{-1}$sr$^{-1}$MeV below a few MeV, increasing to 10$^{-4}$ cm$^{-2}$s$^{-1}$sr$^{-1}$MeV around 10 MeV, and decreasing again to a few 10$^{-5}$ cm$^{-2}$s$^{-1}$sr$^{-1}$MeV above 30 MeV \citep{eastrogamWB}. This is a factor of $\sim$~30~--~10$^3$ below the predicted intensity depending on the energy. 
The most important point is that pion-decay component (red lines) is the major contributor
at energies above $\sim$~100 MeV, while at energies below several ten MeV the IC component (green lines) dominates by far over
any other component. This will allow constraining at best the IC emission, and consequently also the 
bremsstrahlung component (cyan lines). 
As a result, this will also allow to obtain the spatial distribution of CR all-electrons in the Galaxy by studying the bremsstrahlung and the IC separated components.  
Overall, our modeling shows that
observations with e-ASTROGAM and AMEGO {will disentangle} the different interstellar emission mechanisms,
which can not be performed by any current gamma-ray instrument. Besides providing information on CRs, these interstellar components act as confusing background for many
other research topics such as dark matter searches \citep[e.g.][]{P7IG}, source detections \citep[e.g.][]{IEM}, and extragalactic studies \citep[e.g.][]{EGB}. Hence, their better better determination will help in constraining also other components.

 \section{Discussions and Conclusions}
In this work CR propagation models consistent with recent CR measurements are tested against selected available data of the interstellar emission in radio and in gamma rays simultaneously. For the first time, this work shows that this is a feasible approach, which leads to fundamental model constraints, but it also introduces additional challenges. In more detail, we perform this study by comparing propagation models with spectral data of the local gamma-ray HI emissivity and synchrotron observations at intermediate latitudes. 
This approach allows obtaining the all-electron LIS, especially in the range (2 -- 10$^{5}$)~MeV with no assumption on solar modulation. This enables us to test and constrain propagation models. 
Some models consistent with CR measurements only are disfavored, while other models can be put forward. Even though two of our models (PDDE and DRELowV) represent at best the data, we do not find a unique model that can reproduce all the observables at a time.

The main results from this study are: \\
(1) The injection spectral index of primary CR electrons need at least a break below a few GeV. Models with no breaks are excluded because they over-predict {\em Voyager 1}  CR all-electron measurements.  
Our DRC model (diffusion + convection + reacceleration) and our DRE model (diffusion + reacceleration) require two breaks in order to reproduce CR data, while our PDDE model (diffusion only) requires one break only. \\
(2) Models with a high all-electron LIS intensity in the $\sim$(10$^{2}$--10$^{4}$)~MeV range, and hence models that produce a large amount of secondary electrons and positrons, are excluded by both synchrotron and gamma ray observations, even though in agreement with direct CR measurements. This affects reacceleration models with Alfven velocity with the typical value of $\sim$ 30 -- 40 km/s for protons. The consequence is that models with reacceleration need significantly different propagation parameters for low-mass isotope data and for light elements (including secondary-to-primary ratios) in order to be supported by CR measurements, synchrotron and gamma-ray data. On the other hand, the all-electron LIS produced by usual plain diffusion models is supported by CR measurements, and also by synchrotron and gamma-ray data, adopting the same propagation parameters for low-mass isotopes and light elements. We provide our resulting favorite all-electron LIS based on local synchrotron, gamma-ray data, and direct CR measurements. Our finding that some recent propagation models consistent with CR measurements are not supported by multiwavelength observations suggests future propagation parameters studies to be checked against both radio and gamma-ray observations.\\
(3) The calculated spectrum of the local gamma-ray emissivity above  $\sim$1~GeV due to pion decay produced by CRs as precisely measured by AMS-02 is lower than observed, even if accounting for solar modulation. The overall normalization of the proton spectrum derived to fit the emissivity data in the high-energy region free from modulation is $\sim$ 1.3 -- 1.4. This indicates that the direct CR measurements do not represent the average spectrum in the local region within $\sim$1~kpc  probed  by the local gamma-ray emissivity. We provide the normalized proton spectrum that best-fit gamma-ray emissivity data. \\ 

As a general result we identify a preferred propagation model, PDDE, which is a plain diffusion model. This model with enhanced proton spectrum is finally in agreement with synchrotron and gamma-ray data. In Appendix we provide a table with the all-electron and proton spectra for PDDE model. An attempt to identify a model with reacceleration, DRELowV, provides results as good as the PDDE model. \\

We discuss further outcomes driven by this study. \\
(4) For most of the propagation models used in this work (DRE, DRC, DRELowV), by comparing the modeled electron LIS to AMS-02 measurements (Figure~\ref{fig1}) 
it is possible to note that solar modulation on positrons has to be much larger than on electrons. This could imply that positrons have to be modulated differently than electrons in order to fit CR measurements at Earth. 
This might hint the evidence for a charge dependent modulation, and hence the need for a heliospheric propagation scenario more complex than usually assumed. 
However, this charge effect is not evident for models with no reacceleration as the PDDE model. \\
(5) In the Galactic centre region in gamma rays, models having low density of all-electrons in the $\sim$(10$^{2}$--10$^{4}$)~MeV range (i.e. PDDE and DRELowV models) may be favored by {\it Fermi} LAT data at energies below 500 MeV, while in the same energy range DRE and DRC may be disfavored overproducing gamma rays. 
However, for PDDE and DRELowV models, enhancing the ISRF as supported by SPI and COMPTEL data produces many gamma rays in the 100 -- 500 MeV energy band, which may be against observations. This is because an enhanced ISRF would enhance the IC emission at all energies. This may solve the degeneracy between CR all-electrons and ISRF,  supporting the high-energy CR electron origin, and disfavoring the ISRF origin, of the enhanced IC emission found by \cite{P7IG} in the Galactic centre region above 1 GeV. 
However, the Galactic centre is a very complicated region, and it needs further dedicated works in order to finally probe CR density and spectra there. 
The Galaxy is optically thin, hence, when looking at the Galactic centre the interstellar emission acts both as foreground and as background because of the large integration length.
The spatial computations of the gamma-ray emission in this work, as usually done, relay in the 2D azimuthally symmetric distributions of CR sources available in the public version of GALPROP. 
More sophisticated 3D CR source distributions could make some differences in the spatial distribution of the emission, and may be investigated in future studies. 
We have verified that for usual 2D CR source distributions, as used in \cite{diffuse2} and in all the works cited above, the spectrum of the calculated gamma-ray emission is not affected by the assumed CR source distribution. While the Galactic centre is an interesting sky region, it also is a very complicated area
where to draw final conclusions upon.
Moreover, the modeling in this region suffers from large uncertainties given by the gas density along the line of sight. Hence, while we qualitatively discuss the comparison of interstellar models with data, we do not draw any new final conclusion by looking at this region alone, which would need a dedicated and more sophisticated work. In addition, the influence on CRs of Galactic winds and of a possible anisotropy of the diffusion coefficient can be relevant. 
For instance, the possibility to launch CR-induced winds in the Galactic environment has been investigated in very recent works \citep[e.g.][]{Recchia}.
Physical conditions  \citep[e.g.][]{Pfrommer, Giri} can be different from the models used in our work. As an example, recent Galaxy formation simulations \citep[e.g.][]{Pakmor} showed that 
with an anisotropic diffusion most CRs remain in the disk having important consequences for the gas dynamics in the disk, while with an isotropic diffusion CRs are allowed to quickly diffuse out of the disk. This could have unpredictable consequence to our modeling. However, implementing such effects is beyond the present effort.  \\
(6) The X-ray to soft gamma-ray intensity of the diffuse emission in the inner Galaxy observed by SPI and COMPTEL is well reproduced by the IC emission of models having a large all-electron density in the $\sim$(10$^{2}$--10$^{4}$)~MeV range (DRE and DRC models). However these models are in tension with the observed local gamma-ray emissivity and with the observed synchrotron emission. 
This could suggest that SPI and COMPTEL diffuse data in the inner Galaxy region are affected by source contamination of unresolved sources (due to the well-known limited sensitivity and angular resolution of the instruments), which would mimic the IC emission produced by the enhanced all-electron density in the $\sim$(10$^{2}$--10$^{4}$)~MeV range of the DRE and DRC models. Such a possible contaminating source population in the SPI and COMPTEL energy band could be the soft gamma-ray pulsars that were found to have hard power-law spectra in the hard X-ray band and reach maximum luminosities typically in the MeV range \citep{Kuiper}. 
The presence of one or more sources of low-energy CR all-electrons in the inner Galaxy region only could also boost the resulting integrated IC component at X-ray energies, but this would also boost the bremsstrahlung component at few hundred MeV energies, which again would not be supported by {\em Fermi} LAT data in the Galactic center region (see previous point). However, while the inner Galaxy is an interesting sky region, it also is a very complicated area where to draw final conclusions upon. Dedicated analyses and more sensitive observations of the inner Galaxy in the MeV range would be needed in order to have a much clearer picture of this region. \\
(7) In an effort to make predictions for the newly proposed MeV missions, e-ASTROGAM and AMEGO, we have also explored our models at energies below 100 MeV. 
The sky above 100~MeV is dominated by the emission produced by CRs interacting with the gas and the ISRF via pion-decay, IC, and bremsstrahlung. Disentangling the different components at the LAT energies is challenging and is usually done in a model-dependent approach. Uncertainties in the interstellar medium is the major limitation to such a modeling and hence in our knowledge of CRs \cite[e.g.][]{diffuse2}. The situation below 100~MeV is still unexplored. Predictions of present models to such low energies show IC and bremsstrahlung to be the major mechanisms of CR-induced emission, which are of leptonic origin.  
With their improved PSF and energy resolution e-ASTROGAM and/or AMEGO will finally be able to access those energies that have never been studied after the COMPTEL era. \\

The results in this work are important also for future dedicated studies of diffuse emissions in general.
In fact, propagation models, as our PDDE and DRELowV models, could be used as baseline models to assess uncertainties of the gamma-ray interstellar emission in the entire sky, 
for example for studies regarding diffuse gamma-ray emissions \citep[e.g.][]{P7IG} and extended sources \citep[e.g.][]{SNR}.

\section*{Acknowledgements} 
E. Orlando acknowledges support from NASA Grants No. NNX16AF27G. \\
The author thanks the referee for substantial comments that have improved the manuscript.
Useful discussions with Gudlaugur J\'ohannesson, Igor Moskalenko and Andy Strong are acknowledged. \\
This work makes  use of HEALPix\footnote{http://healpix.jpl.nasa.gov/} described in \cite{healpix}.








\newpage

\section*{Appendix} 

We report the LIS of all-electrons and protons for the favorite model PDDE in Table~\ref{TableApp1} and Table~\ref{TableApp2} respectively. The proton spectrum scaled to fit the local gamma-ray emissivity is reported in Table~\ref{TableApp3}.  \\

\tablecaption{All-electron spectrum for PDDE model that fits CR measurements, synchrotron emission and gamma ray emissivity, as plotted in Figure~\ref{fig1}. The first column is the kinetic energy and the second column is the spectral intensity multiplied by E$^2$.}
\tablefirsthead{\toprule E&\multicolumn{1}{c}{Intensity} \\
(MeV)&\multicolumn{1}{c}{(MeV$^2$cm$^{-2}$s$^{-1}$sr$^{-1}$Mev$^{-1}$)}
\\ \midrule}
\center
\tablehead{%
\multicolumn{2}{c}%
{{\bfseries  Continued from previous column}} \\
\toprule
 E &\multicolumn{1}{c}{Intensity}\\ 
 (MeV)&\multicolumn{1}{c}{(MeV$^2$cm$^{-2}$s$^{-1}$sr$^{-1}$Mev$^{-1}$)}
 \\ \midrule}
\tabletail{%
\midrule \multicolumn{2}{r}{{Continued on next column}} \\ \midrule}
\tablelasttail{%
\\\midrule
\multicolumn{2}{r}{{Concluded}} \\ \bottomrule}
\begin{supertabular}{ll}
2.9243 & 0.948432171 \\
3.4969 & 1.09652327 \\
4.1817 & 1.263044372 \\
5.0006 & 1.450084145 \\
5.9799 & 1.6601582 \\
7.151 & 1.89549382 \\
8.5513 & 2.15923726 \\
10.226 & 2.45476391 \\
12.228 & 2.7857931 \\
14.623 & 3.15681 \\
17.487 & 3.5728002 \\
20.911 & 4.0395083 \\
25.006 & 4.563631 \\
29.903 & 5.153056 \\
35.759 & 5.81666 \\
42.762 & 6.562885 \\
51.136 & 7.389989 \\
61.15 & 8.23707 \\
73.125 & 8.67137 \\
87.445 & 8.93157 \\
104.57 & 9.20788 \\
125.05 & 9.49295 \\
149.54 & 9.77528 \\
178.82 & 10.04511 \\
213.84 & 10.30112 \\
255.71 & 10.54979 \\
305.79 & 10.79938 \\
365.67 & 11.05006 \\
437.28 & 11.29157 \\
522.92 & 11.51173 \\
625.32 & 11.69561 \\
747.77 & 11.82816 \\
894.21 & 11.89636 \\
1069.3 & 11.88194 \\
1278.7 & 11.76445 \\
1529.1 & 11.51491 \\
1828.6 & 11.10544 \\
2186.7 & 10.51503 \\
2614.9 & 9.74485 \\
3127.0 & 8.81966 \\
3739.3 & 7.78374 \\
4471.6 & 6.67781 \\
5347.3 & 5.48447 \\
6394.5 & 4.39294 \\
7646.7 & 3.51204 \\
9144.1 & 2.80351 \\
10935.0 & 2.235134 \\
13076.0 & 1.780259 \\
15637.0 & 1.416798 \\
18699.0 & 1.126881 \\
22361.0 & 0.895903 \\
26740.0 & 0.711662 \\
31976.0 & 0.565024 \\
38238.0 & 0.448214 \\
45727.0 & 0.36064 \\
54681.0 & 0.2822996 \\
65389.0 & 0.2236471 \\
78195.0 & 0.1770424 \\
93507.0 & 0.1400206 \\
111820.0 & 0.1106436 \\
133720.0 & 0.0873381 \\
159900.0 & 0.0688765 \\
191220.0 & 0.0542632 \\
228660.0 & 0.04271027 \\
273440.0 & 0.03358651 \\
326990.0 & 0.02638933 \\
391020.0 & 0.02071834 \\
467600.0 & 0.0162543 \\
559170.0 & 0.01274445 
\end{supertabular}%
\label{TableApp1}

\tablecaption{Proton spectrum for baseline PDDE model that fits CR measurements, but it does not fit the local emissivity, as plotted in Figure~\ref{fig1a}. The first column is the kinetic energy and the second column is the spectral intensity multiplied by E$^2$.}
\tablefirsthead{\toprule E&\multicolumn{1}{c}{Intensity} \\
(MeV)&\multicolumn{1}{c}{(MeV$^2$cm$^{-2}$s$^{-1}$sr$^{-1}$Mev$^{-1}$)}
\\ \midrule}
\center
\tablehead{%
\multicolumn{2}{c}%
{{\bfseries  Continued from previous column}} \\
\toprule
 E &\multicolumn{1}{c}{Intensity}\\ 
 (MeV)&\multicolumn{1}{c}{(MeV$^2$cm$^{-2}$s$^{-1}$sr$^{-1}$Mev$^{-1}$)}
 \\ \midrule}
\tabletail{%
\midrule \multicolumn{2}{r}{{Continued on next column}} \\ \midrule}
\tablelasttail{%
\\\midrule
\multicolumn{2}{r}{{Concluded}} \\ \bottomrule}
\begin{supertabular}{ll}
29.903 & 2.389 \\
35.759 & 3.4093 \\
42.762 & 4.8157 \\
51.136 & 6.7323 \\
61.15 & 9.3142 \\
73.125 & 12.752 \\
87.445 & 17.285 \\
104.57 & 23.211 \\
125.05 & 30.913 \\
149.54 & 40.877 \\
178.82 & 53.733 \\
213.84 & 70.31 \\
255.71 & 91.668 \\
305.79 & 119.1 \\
365.67 & 154.46 \\
437.28 & 199.75 \\
522.92 & 256.4 \\
625.32 & 318.41 \\
747.77 & 335.54 \\
894.21 & 354.13 \\
1069.3 & 373.29 \\
1278.7 & 391.29 \\
1529.1 & 407.45 \\
1828.6 & 421.3 \\
2186.7 & 432.31 \\
2614.9 & 440.01 \\
3127.0 & 443.85 \\
3739.3 & 442.27 \\
4471.6 & 396.72 \\
5347.3 & 333.52 \\
6394.5 & 299.64 \\
7646.7 & 275.33 \\
9144.1 & 251.15 \\
10935.0 & 227.61 \\
13076.0 & 205.08 \\
15637.0 & 183.82 \\
18699.0 & 164.02 \\
22361.0 & 145.75 \\
26740.0 & 129.05 \\
31976.0 & 113.93 \\
38238.0 & 100.31 \\
45727.0 & 88.117 \\
54681.0 & 77.257 \\
65389.0 & 67.623 \\
78195.0 & 59.106 \\
93507.0 & 51.599 \\
111820.0 & 44.996 \\
133720.0 & 39.202 \\
159900.0 & 34.128 \\
191220.0 & 29.692 \\
228660.0 & 25.818 \\
273440.0 & 22.438 \\
326990.0 & 19.493 \\
391020.0 & 16.929 \\
467600.0 & 14.697 \\
559170.0 & 12.757 \\
668670.0 & 11.07 \\
799610.0 & 9.6046 \\
956200.0 & 8.3318 \\
\end{supertabular}%
\label{TableApp2}


\begin{table}
\begin{center}
\caption{Proton spectrum obtained from the emissivity, as plotted in Figure~\ref{fig4}. The first column is the kinetic energy and the second column is the spectral intensity multiplied by E$^2$.}
\begin{tabular}{cc}
\\
 \hline
 \hline
 {E} &  {Intensity}          \\
 {(MeV)} &  {(MeV$^2$cm$^{-2}$s$^{-1}$sr$^{-1}$Mev$^{-1}$)} 
  \\
 \hline
894.21 & 460.369 \\
1069.3 & 485.277 \\
1278.7 & 508.677 \\
1529.1 & 529.685 \\
1828.6 & 547.69 \\
2186.7 & 562.003 \\
2614.9 & 572.013 \\
3127.0 & 577.005 \\
3739.3 & 574.951 \\
4471.6 & 515.736 \\
5347.3 & 433.576 \\
6394.5 & 389.532 \\
7646.7 & 357.929 \\
9144.1 & 326.495 \\
10935.0 & 295.893 \\
\hline
\\
\label{TableApp3}
\end{tabular}
\end{center}
\end{table}


\bsp	
\label{lastpage}
\end{document}